\documentclass{emulateapj}
\usepackage{epsfig}

\newcommand{\beq}{\begin{equation}}
\newcommand{\eeq}{\end{equation}}

\def\be{\begin{equation}}
\def\ee{\end{equation}}
\def\bea{\begin{eqnarray}}
\def\eea{\end{eqnarray}}



\def \logTd6 {\hbox{log$( T/6 \kev)$} }

\def\myputfigure#1#2#3#4#5%
{\vskip#5pt\makebox[0pt]{\hskip#2in
\includegraphics[width=#3\textwidth]{#1}}\vskip#4pt\hfill}

\def \arcmin     { ^{\prime} }
\def \arcsec    {^{\prime\prime}}

\def \kms            {~{\rm km~s}^{-1}}

\def \etal      {et al.}

\def \kmsmpc    {{\rm\ km\ s^{-1}\ Mpc^{-1}}}
\def \kev       {{\rm\ keV}}

\def \hmsol     {h^{-1}{\rm\ M}_\odot}
\def \hMpc      {h^{-1}{\rm\ Mpc}}
\def \hkpc      {h^{-1}{\rm\ kpc}}

\input epsf
\bibpunct{(}{)}{;}{a}{}{,}

\begin{document}

\lefthead{CLUSTER LENSES}
\righthead{Hennawi et al.}

\title{Characterizing the Cluster Lens Population}

\author{Joseph F. Hennawi\altaffilmark{1,2,3}, 
  Neal Dalal\altaffilmark{1,4}
  Paul Bode\altaffilmark{3}
  Jeremiah P. Ostriker\altaffilmark{3}
} 

\altaffiltext{1}{Hubble Fellow}

\altaffiltext{2}{Department of Astronomy, University of California
  Berkeley, Berkeley, CA 94720}

\altaffiltext{3}{Princeton University Observatory, Princeton, NJ 08544}

\altaffiltext{4}{Institute for Advanced Study, Einstein Drive, Princeton, NJ 08540 }

\begin{abstract}
  We present a detailed investigation into which properties of cold dark
  matter halos make them effective strong gravitational lenses. The
  cross sections for giant arc formation of 878 clusters from a
  high-resolution N-body simulation of the $\Lambda$CDM cosmology are
  measured by ray tracing through 13,594 unique projections. We
  measure concentrations, axis ratios, orientations, and the amount of
  substructure of each cluster, and compare the lensing weighted
  distribution of each quantity to that of the cluster population as a
  whole.  We find that NFW profiles provide just as good a fit to
  lensing clusters as they do to the total cluster population;
  however, the concentrations of lensing clusters are on average
  $34\%$ larger than the typical cluster in the Universe. Despite this
  bias, the anomalously high concentrations of lensing clusters ($c >
  14$) recently measured by several groups from combined strong and
  weak lensing analyses \citep{Kneib03,Gavazzi03,Broad05b}, appear to
  be inconsistent with the concentration distribution in our
  simulations, which predict $< 2\%$ of lensing clusters should have
  concentrations this high.  No correlation is found between strong
  lensing cross section and the amount of substructure, indicating
  that the population of cluster lenses is no more relaxed or
  disturbed than typical clusters in the Universe.  Lensing clusters
  tend to have their principal axis aligned with the the line of
  sight: the median angle is $\left|\cos\theta\right|=0.67$.  We
  introduce several different types of simplified dark matter halos,
  and use them to isolate which properties of CDM clusters make them
  effective gravitational lenses. Projections of halo substructure
  onto small radii and the large scale mass distribution of clusters
  do not significantly influence strong lensing cross sections. The
  abundance of giant arcs is primarily determined by the mass
  distribution within an average overdensity of $\sim 10,000$. The clumpy cores
  of dark matter halos result in $\sim 25-60\%$ more giant arcs than
  smooth ellipsoids of the same total mass. A multiple lens plane ray
  tracing algorithm is used to show that projections of large scale
  structure increase strong lensing cross sections by a 
  modest amount $\lesssim 7\%$.  We revisit the question of whether
  there is an excess of giant arcs detected for high redshift clusters
  in the Red-Sequence Cluster Survey \citep{Glad03} and find that the
  number of high redshift ($z\gtrsim 0.6$) lensing clusters is in good
  agreement with $\Lambda$CDM, although our simulations predict 
  more low redshift ($z\lesssim 0.6$) lensing clusters than were observed. 
\end{abstract}


\section{Introduction}
\label{sec:intro}

Strong gravitational lensing by clusters of galaxies provides a unique
laboratory for studying the small scale dark matter distribution of
the largest collapsed structures in the Universe.  Because strong
lensing directly probes the gravitational potential, it is free from
the assumptions which plague other techniques which probe the mass
distribution on comparable scales. For example, dynamical measures
\citep{NK96,Kelson02} assume that dark halos are virialized systems
and models of the X-ray temperature profile of the intracluster medium
\citep[e.g.][]{ABG02,Ettori02,LBS03} typically assume hydrostatic
equilibrium.

Upcoming X-ray \citep{Romer01,Ebel01} Sunyaev-Zeldovich
\citep{Carl02,Koso03,Schwan03}, and optical \citep{GY04} cluster searches will
dramatically increase the number of clusters of galaxies known. Deep
follow up imaging of these clusters will discover hundreds of new
giant arcs. Such statistical samples of arcs will allow measurement of
the giant arc abundance as a function of cluster redshift, the radial
distributions of arcs from the cluster center, and the distribution of
relative angles between arcs in multi-arc systems.  Comparisons of
these quantities to expectations from ray tracing simulations of
clusters in a $\Lambda$CDM cosmology have already been carried out
\citep{Bart98,Wamb04a,DHH,Shirley04,Li05} using small samples of $\sim 20$
known arcs \citep[see][for a recent compilation]{Sand05}.

Armed with future statistical samples, we will be able to measure the
distributions of cluster lens properties, in addition to the simple
`one point' statistics of the arcs. Detailed modeling of image
positions in cluster lenses can determine a set of best fit parameters
which describe the distribution of dark matter in each cluster lens
\citep[e.g.][]{Tyson98,Smith01,Sand04,Broad05a}, and even stronger 
constraints can be obtained when strong lensing information is combined with
larger scale weak lensing measurements
\citep{Kneib03,Gavazzi03,Broad05b}. These parameters might include the
slope of the mass profile, its concentration, or the projected
ellipticity of the cluster.  A comparison of the observed distribution
of these parameters to the expected structures from cosmological
simulations of a $\Lambda$CDM Universe will provide a strict test of
our current theory of dark matter and structure formation.

In this context, the questions naturally arise: Do strong
lensing selected clusters constitute a fair sample of clusters in the
Universe? Do biases exist with respect to halo concentration,
triaxiality, preferential alignment, or the amount mass in
substructure?  Clearly, these biases must be taken into account before
a comparison can be made between the observed distribution of cluster
lens properties and the properties of numerically simulated clusters.
Additionally, by comparing the distribution of lensing cluster
properties to the cluster population as a whole, we can hope to gain
insights and intuition about what makes a CDM cluster an effective
strong lens.

In this paper we statistically characterize the properties of the
cluster lens population.  Using the ray tracing technique discussed
\citet[][henceforth DHH]{DHH}, we compute the lensing cross sections
of a large sample of clusters from a cosmological N-body
simulation. We measure the properties of each cluster and compare the
lensing weighted distribution of cluster properties to the cluster
population as a whole. We also attempt to isolate which properties of
CDM clusters make them effective gravitational lenses. In this context
we introduce `Analog Halos,' which are simplified dark matter halos
which retain one or more of the properties of the original simulated
clusters.  As an aside, we also quantify the effect of projections of
large scale structure between the source plane and the observer on the
abundance of giant arcs. Armed with lensing cross sections for
clusters in a large simulation volume, we revisit the question of
whether there is an excess of giant arcs detected for high redshift
clusters in the Red-Sequence Cluster Survey (RCS) \citep{Glad03}.  We
were unable to answer this question definitively in DHH because the
cosmological volume simulated in that study was too small \citep[see
  also][]{Wamb04a}.

The cosmological N-body simulation and our methodology for measuring
the cluster parameters and quantifying substructure are described in
\S~\ref{sec:sims}.  We review the ray tracing algorithm in
\S~\ref{sec:ray_trace} and investigate the effect of additional matter
in the light cone in \S~\ref{sec:light_cone}.  In \S~\ref{sec:analog},
we introduce the `Analog Halos' and examine which properties of CDM
halos make them effective gravitational lenses.  We compare the
statistical properties of cluster lenses to the total cluster
population in \S~\ref{sec:histo} . The abundance of giant arcs in the
RCS is revisited in \S~\ref{sec:RCS}, and we summarize and conclude in
\S~\ref{sec:conc}.

\section{Simulated Clusters}
\label{sec:sims}

In this section, we describe the simulated clusters which are the
input to our ray tracing algorithm. For the analysis in
\S~\ref{sec:analog} and \S~\ref{sec:histo} we will require various
properties of clusters, such as triaxiality, the amount of
substructure, and cluster concentration both in three dimensions and
for two dimensional projections.  First we describe the N-body
simulations and how the clusters were identified, then we describe how
cluster properties were measured.

\subsection{N-body simulations}

\label{sec:Nbody}

In order to simulate the strong lensing effects of a Universe filled
with dark matter, we used clusters drawn from a cosmological N-body
simulation.  The simulation was performed with the Tree-Particle-Mesh
(TPM) code of \citet{BO03}.  TPM uses the Particle-Mesh method for
long-range forces and a tree code for sub-grid resolution; individual
isolated, overdense regions are each treated as a separate tree, thus
ensuring efficient use of parallel computers.  The following
cosmological parameters were used: matter content $\Omega_{\mathrm
  M}=0.3$, cosmological constant $\Omega_\Lambda=0.7$, Hubble constant
$H_0=70~\kmsmpc$, linear amplitude of mass fluctuations
$\sigma_8$=0.95, and primordial power spectral index $n_s$=1.  These
parameters are consistent (within 1$\sigma$) of the WMAP derived
cosmological parameters \citep{Spergel03}.  The simulation volume is a
periodic cube with comoving side length of $L=320~\hMpc$ containing
$N=1024^3$ particles, so the particle mass is $m_{\rm p}=2.54\times
10^9 \hmsol$.  The cubic spline softening length was set to
$\epsilon=3.2 \hkpc$. The small softening length, small particle mass,
and large volume make this simulation ideal for studying strong
lensing by clusters; this simulation was previously analyzed by
\citet{Wamb04a,Wamb04b}.  We used outputs at seven different `snapshots',
spaced a comoving length $L$ apart and covering the range of redshifts
over which the critical density is low enough to produce an
appreciable amount of strong lensing. These redshifts are
$z=0.17,0.29,0.41,0.55,0.70,0.87,1.05,1.26,$ and $1.49$.

A ``friends--of--friends'' (FOF) group finder \citep{Davis85} with the
canonical linking length of $b=0.2$ was applied to each particle
distribution to identify cluster size dark matter halos. For each
cluster with a FOF group mass above $M_{\rm FOF}\geq 10^{14}\hmsol$,
all the particles within a $5~\hMpc$ sphere about the center of mass
were dumped to separate files and used as inputs to our ray tracing
code.

\subsection{NFW Profile Parameters}
\label{sec:NFW}

\begin{figure*}
  \centerline{
    \raisebox{0.4cm}{\epsfig{file=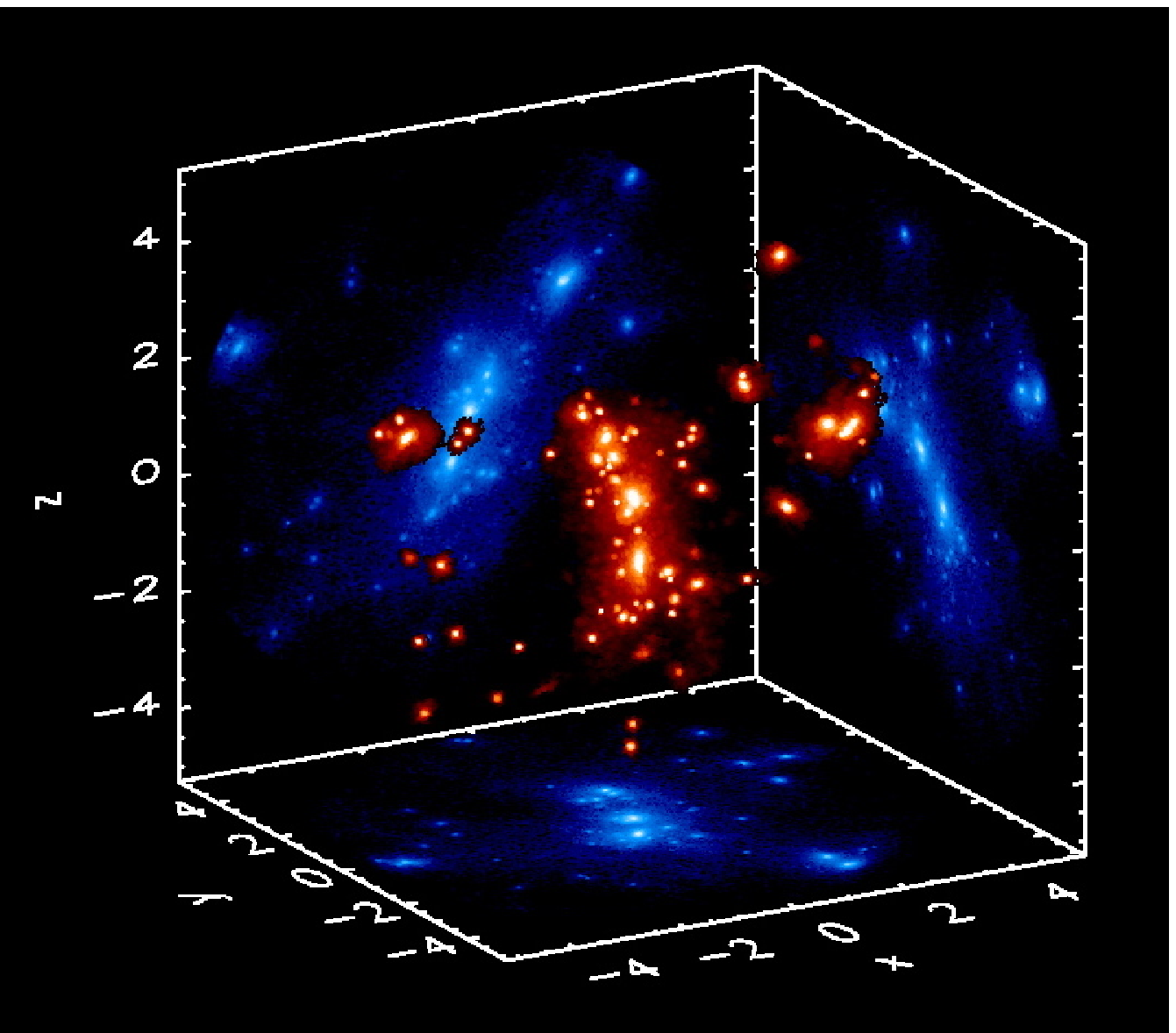,width=0.50\textwidth}}
    \epsfig{file=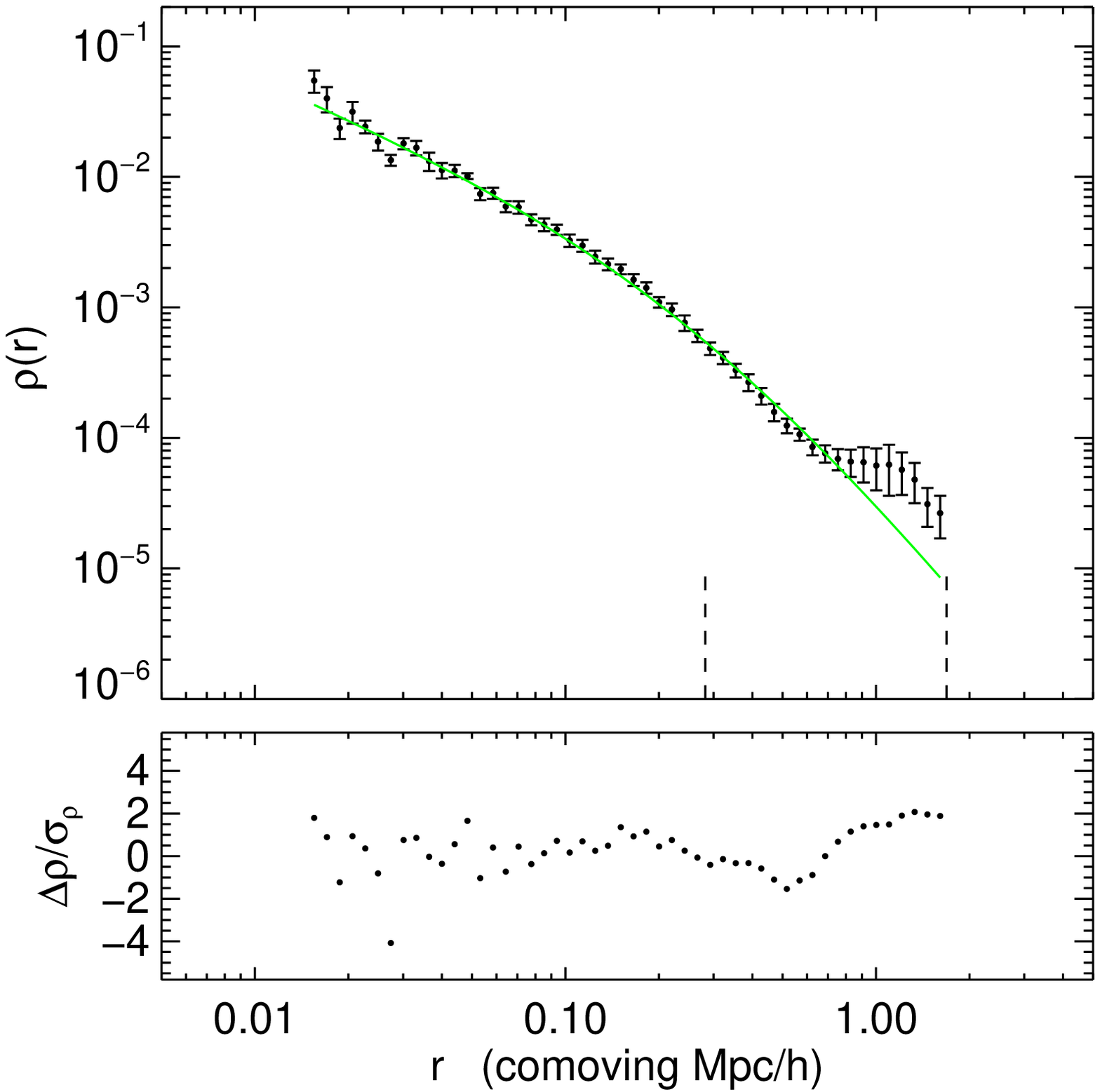,width=0.50\textwidth}}
  \caption{ Example of a cluster with a large amount of substructure
    inside the virial radius. \emph{Left:} Three dimensional density
    distribution of the cluster. Surface density projections on each
    coordinate plane are also shown to aid visualization. The spatial
    axes are in units of comoving $\hMpc$ and both density (2-d and
    3-d) color maps are logarithmically stretched. \emph{Right:}
    Radial density profile (points) and best fit NFW profile fit
    (solid green line), which has a reduced $\chi^2=1.28$. The y-axis
    has units of $\hmsol/\rm {pc}^3$. The inner and outer dashed
    vertical lines indicate the locations of the scale radius and
    virial radius for the NFW fit, respectively. The lower panel plots
    the ratio of $\Delta\rho$, which is the difference between the
    profile and the fit, to $\sigma_{\rho}$, the bootstrapped density
    variance, for each radial bin. The bootstrapped variances properly
    down weight the bump in the outer radial bins caused by the
    massive substructures.  Had we used the Poisson errors alone the
    radial bump would have had a catastrophic effect on the fit
    because $1\slash\sqrt{N}$ is very small for these outer bins.
    \label{fig:NFW_fit}
  }
\end{figure*}

For each cluster identified by FOF, the gravitational potential was
computed and the center of the cluster was defined to be the most
bound particle in the $5~\hMpc$ particle dump. The radial mass profile
of the cluster about this center was calculated, and the virial radius
was defined as the radius of the innermost particle at which the
average density interior to it was greater than or equal to
$\Delta_{\rm vir}(z)$ times the cosmic \emph{mean} matter density
where the virial overdensity is $\Delta_{\rm vir} \approx
\frac{18\pi^2 + 82 x-39 x^2}{1+x}$ \citep{BN98}, with $x\equiv
\Omega_{m}(z) - 1$.

We computed the number of particles in 50 logarithmically spaced
radial bins between the 100th particle from the cluster center and
the virial radius.  The smallest radial bin was placed at the position
of the 100th particle because this roughly defines the scale at which the mass
distribution is unaffected by two body relaxation effects
\citep{Power03}. This scale exceeds the softening length for all of the 
clusters we considered and provides a stricter limit on the scales we 
resolved. 

We then fit the number of particles in each bin $dN_{i}$ to the
\emph{number profile} corresponding to the universal density profile
of \citet{NFW} (henceforth NFW). The number of particles in each bin 
$dN_{i}\sim r^2\rho dr$ is much less noisy than the density and furthermore, 
it is $dN_{i}$ which should obey Poisson statistics if counting errors
are the only source of noise in each measurement, allowing us 
to assign sensible errors (see discussion below).  

Thus we fit the quantity
\be 
dN_i(r_{\rm s},M_{\rm C})=M_{\rm C} \left[A(x_{i})-A(x_{i-1})\right]  
\label{eqn:NNFW}
\ee 
for the two parameters  $M_{\rm C}$ and $r_{\rm s}$, 
where
\be
A(x)\equiv \ln(1 + x) -\frac{x}{1+x}.
\ee 
and $x=r/r_{\rm s}$. In terms of the familiar NFW parameters
$M_{\rm C} = 4\pi\rho_{\rm s} r^3_{\rm s}\slash m_{\rm p}$, but we opt to fit for $M_{\rm C}$ 
because the profile is linear in this quantity.

The parameters are determined by minimizing the reduced $\chi^2$ 
penalty function 
\be 
\chi^2 = \frac{1}{50}\sum_{i=1}^{50} \frac{\left[dN_i-dN_i(r_{\rm
      s},M_{\rm C})\right]^2}{\sigma^2_i}.  \label{eqn:chi2}
\ee

Our computation of the variance for each logarithmic bin, $\sigma^2_i$,
warrants further discussion. The deviations between our simulated
clusters and an NFW profile can be caused by both counting errors, 
described by Poisson statistics, and massive substructures or
asymmetries in the cluster.  Without taking the latter source of error
into account, a bump in the radial profile of a halo will have a
catastrophic effect on the NFW fit if it is in the outer region of the
halo, because the formal Poisson errors for these radial bins will be very
small. Previous workers have dealt with this problem by removing these
substructures from the halo \citep[e.g.][]{JS02} or truncating the
profile and the fit wherever these bumps occur
\citep[e.g.][]{Bullock01}.  However, in the context of strong lensing,
neither of these approaches is appropriate.  We wish to measure a
three dimensional profile which is as closely related as possible to
the two dimensional surface density profile probed by strong lensing
observations. Strong lensing is sensitive to the total projected mass,
and since it is not possible to remove or truncate substructures
observationally we opt not to do so in the simulations either.

Instead, we define a modified variance which incorporates halo
asymmetry but which reduces to the Poisson variance for a spherically
symmetric halo. Each radial bin is divided into 12 equal area sectors
using Healpix\footnote{http://www.eso.org/science/healpix},  and we 
estimate  the variance via a `bootstrap' 
\be
\sigma^2_i\equiv \frac{n_{\rm pix}}{n_{\rm pix}-1}\sum_{j=1}^{n_{\rm
    pix}}\left(dn_{ij} -\frac{dN_i}{n_{\rm pix}}\right)^2 , \label{eqn:signfw}
\ee 
where $n_{\rm pix}=12$ is the number of Healpix sectors and $dn_{ij}$ is the
number of particles in the $i$th radial bin which land in angular
sector $j$.

The right panel of Figure \ref{fig:NFW_fit} compares the measured
density profile to the best fit NFW profile\footnote{Although we fit
  the number profiles using $dN_i$ and $\sigma_i$, we plot density
  profiles $\rho(r)$ and corresponding propagated errors
  $\sigma_{\rho}$ for the sake of visualization.}, for a cluster from
our simulation which is depicted in the left panel. The massive
substructure interior to the virial radius causes a bump in the radial
density profile, however, the bootstrapped errors correctly down
weight these points and minimize their influence on the fit.

The distribution of the reduced $\chi^2$ for the 878 clusters
in our fiducial snapshot is displayed in Figure \ref{fig:chi2_hist}.
The fact that median value of this distribution is near unity gives us
confidence that the NFW profile is providing a good fit to our
simulated clusters and our bootstrapped $\sigma_i$ are accurately
describing the Poisson errors as well as the deviations from symmetry.

Once $M_{\rm C}$ and $r_{\rm s}$ are determined, all the NFW
parameters $r_{\rm s}$, $\rho_{\rm s}$, $c_{\rm vir}$, $r_{\rm vir}$
and $M_{\rm vir}$ can be determined from simple algebraic
relations. Although the NFW fit determines the virial mass of the
cluster, we instead adopt the non-parametric virial mass given by the
mass interior to the virial radius defined by the overdensity
criterion $\Delta_{\rm vir}$.

We also perform fits of the surface mass density to a two dimensional
projected NFW profile, for various projections through the mass
distribution of each cluster. These two dimensional fits are of
interest because the projected mass distribution is the quantity
accessible to observations of strong lensing.  We repeat the fitting
procedure described above for surface mass density by computing the
azimuthally averaged radial profile in 50 logarithmically spaced bins
between 0.02 and 2~$\hMpc$.  The same bootstrapping procedure
(eqn.~\ref{eqn:signfw}) is used to compute the errors, where we use 20
azimuthal bins to estimate the error for each radial bin.  We minimize
the $\chi^2$ of a fit to an analytical expression for the projected
NFW profile \citep{Bart96,WB00}.  Note that this formula for the
projected NFW profile assumes the profile extends to infinity, which
results in a slight bias of 6\% (see \S~\ref{sec:conc2d}) in the
concentrations measured. For the sake of simplicity, we use this
analytical form rather than fitting to the exact numerical expression
for a projected truncated NFW. This small bias does not have a
significant effect on our conclusions.

\subsection{Substructure}
\label{sec:sub}

Although complicated algorithms exist to identify substructures in
N-body simulations \citep[see e.g.][]{Krav03,DL04,Weller04,GKG04}, for
our purposes we are only interested in identifying substructures
massive enough to influence the strong lensing cross sections of
massive clusters, for which even a very simple procedure will
suffice. In addition, there is no need to use velocity information to
unbind particles as is done in other algorithms since the energy of
the matter distribution is irrelevant to its lensing effect.  To this
end, we identify substructures by rerunning the FOF algorithm on the
particles in each cluster dump, but with a smaller linking linking
length of $b=0.05$. 
This causes a cluster halo to fragment into one or
more dense subhalos. The most massive subhalo will be
centered on the potential minimum or center of the cluster, and we
designate this as the cluster core.  We define as a substructure any
additional subhalos (besides the most massive) identified by the FOF
algorithm with a mass $M > 10^{12}\ \hmsol$. This is safely larger
than the smallest subhalo resolved by our particle mass $\sim 100
m_{p}=2.5\times 10^{11}~\hmsol$.

A linking length $b=0.05$ groups particles within an isodensity
contour of $\sim 4,000$ times the background density \citep{LC94}. For
an NFW cluster, the smaller linking length truncates the cluster at a
radius $\sim r_{10,000}$, where the average overdensity is $\sim
10,000$ times the mean density. For, $M_{\rm vir}\sim 5\times
10^{14}~\hmsol$, $c_{\rm vir}=6$, and $z=0.4$, $r_{10,000}\sim
300~\hkpc$ (comoving), which projects to an angular separation of
$\sim 1\arcmin$.


%
%

We define three different statistics which quantify the amount of
substructure in a cluster. Two of the three can be written as 
\be
M_{\rm sub} \equiv \frac{1}{M_{\rm norm}}\sum_{i=2}^{N_{\rm sub}} M_i,
\label{eqn:msub}
\ee where $M_i$ refers to the mass of the $i^{\rm th}$ subhalo,
$M_{\rm norm}$ is a normalizing mass, and the sum extends over all
substructures enclosed within the virial radius of the halo.  Note
that the sum excludes subhalo $i=1$, which as noted above corresponds
to the cluster core.

We explore two different normalization masses $M_{\rm norm}=M_{\rm
  core}$ and $M_{\rm norm}=M_{\rm vir}$, corresponding to the two
statistics $M_{\rm sub-core}$ and $M_{\rm sub-vir}$, respectively. The
statistic $M_{\rm sub-core}$ is normalized by the mass of the dense
core which is the mass relevant to lensing.  This could be biased low
for lensing clusters since they might have more massive cores which is
why we also consider $M_{\rm sub-vir}$. Our third substructure
statistic is $M_{\rm sub-2}=M_2/M_{\rm vir}$, where $M_2$ is the second most
massive sub-halo. This statistic provides a measure of the binarity of the
cluster, and is less sensitive to lower mass halos which may dominate the
total mass in substructure because they are more abundant. 


\subsection{Triaxiality}
\label{sec:tri}
Previous work has shown that for CDM halos, the degree of triaxiality
increases (i.e., the axis ratios decrease) toward the center
\citep{BE87,Warren92,JS02,Schulz05}. For our purposes, we are interested in the
triaxiality of the mass within the dense core which is 
primarily responsible for
the strong lensing.  As mentioned in the previous section, we identify
the most massive subhalo selected by the FOF algorithm using a smaller
linking length with the cluster core.  We compute the principal axes 
and axis ratios by diagonalizing the normalized moment of inertia
tensor for the particles which are linked into this core subhalo
\be
{\bf Q}=\frac{1}{N_{\rm core}}\sum_{i}^{N_{\rm core}} {\bf r}_i
\otimes {\bf r}_i,
\label{eqn:Q}
\ee 
where $N_{\rm core}$ is the number of particles grouped into the
core subhalo.  If $a^2 > b^2 > c^2$ are the principal components of
${\bf Q}$, we compute the axis ratios $q_2\equiv b/a$ and $q_3 \equiv
c/a$ for each cluster.  The eigenvectors of ${\bf Q}$ give the
orientation of the principal axes ${\bf V_a}$, ${\bf V_b}$, ${\bf
  V_c}$. Note that by restricting attention only to particles which
lie in the FOF group, we are computing the axis ratios of the particles 
within an isodensity contour, similar to the approach of \citet{JS02}. 
Other workers have computed the axis ratios of particles
within an ellipsoid \citep{BE87,Warren92} defined by an iterative procedure, 
but these iterations are not guaranteed to converge \citep{JS02,Schulz05} 
which is why opt for the method used here. 

\section{Strong Lensing Calculation}
\label{sec:ray_trace}



We compute the strong lensing properties of our simulated clusters 
using ray-tracing; cross sections and arc statistics are
determined via Monte Carlo integration over the background source plane.
We refer the reader to DHH for details of the ray tracing and the
Monte Carlo calculation.
The ray trace and Monte Carlo are repeated for each projection
through the cluster and for each source plane considered. We ray trace
five source planes at $z_{\rm s}=1.0$, $1.5$, $2.0$, $3.0$, and $4.0$.
The number counts of background galaxies are collapsed into five
source redshift bins centered on these source planes given by
the intervals
$[0.75,1.25]$, $[1.25,1.75]$, $[1.75,2.5]$, $[2.5,3.5]$, and
$[3.5,6.0]$, respectively. Because the critical density for strong
lensing is a slowly varying function of source redshift, this binning
should not introduce significant errors in our calculation.  Given the
cross sections for each source plane and the number counts of background 
galaxies in each bin, the number of giant arcs behind each cluster 
orientation can be computed as a function of limiting magnitude 
or surface brightness. 

DHH and \citet{Shirley04} found that the lensing cross section for a
single cluster varies enormously as a function of orientation (see
Figure~3 of DHH). A factor of $\sim 20$ spread in lensing cross
section was not uncommon and the histogram of cross sections for
different orientations was quite skewed.  To compute reliable mean
cross sections for a cluster, we must thus average over many
projections to appropriately sample the distribution of cross
sections. This orientation averaging is most important for the rarest,
most massive clusters in the simulation volume that dominate the total 
lensing optical depth.  We computed mean lensing cross
sections by averaging over 125 orientations for all clusters with
$M_{\rm FOF}\ge 10^{14.7}~\hmsol$, 31 orientations for clusters in the
mass interval $10^{14.3}~\hmsol\le M_{\rm FOF}<10^{14.7}~\hmsol$, and 3
orientations for all clusters in the range $10^{14}~\hmsol \le M_{\rm
  FOF}<10^{14.3}~\hmsol$.


Given the cross sections for each cluster orientation, the total number of
arcs caused by the clusters in the snapshot at redshift $z_j$ 
\bea
N_{j}(>\theta)=\sum_{k}\Omega n_k\frac{V_j}{L^3}\sum_{i}^{N_{\rm
    clusters}}{\bar \sigma}_{ijk} (>\theta).\label{eqn:Nj}
\eea
Here ${\bar \sigma}_{ijk} (>\theta)$ is the orientation averaged cross
section for forming giant arcs with angular separation larger than
$\theta$ for the $i$th cluster in snapshot $z_j$ with source plane
$z_k$; $\Omega$ is the solid angle of the survey under consideration, $L^3$
is simulation volume, and $n_k$ is the number density of background
galaxies in the bin about source plane $z_k$. The quantity $V_j$ is
the effective volume represented by the snapshot 
\be V_j\equiv
\int_{\frac{z_{j} + z_{j-1}}{2}}^{\frac{z_j +
    z_{j+1}}{2}}\frac{dV}{d\Omega dz}dz, \label{eqn:Vj} 
\ee 
where $\frac{dV}{d\Omega dz}$ is the cosmological volume element. Hence, we
take the clusters in the snapshot at redshift $z_j$ to be representative
of the lensing rate in the volume of the light cone over the redshift
interval $[\frac{z_{j} + z_{j-1}}{2},\frac{z_j + z_{j+1}}{2}]$.

Finally, in what follows we will often use the effective cross section
of a cluster, which we define as
\be
\sigma_{\rm eff}(>\theta)\equiv 
\frac{\sum_{k}n_{k}{\bar \sigma}_{k} (>\theta)}{\sum_{k}n_{k}}. 
\label{eqn:sigeff}
\ee

We defer a discussion of the counts of background galaxies in each
source redshift bin, $n_k$, to \S~\ref{sec:gals}.

\subsection{Adding Brightest Cluster Galaxies}
\label{sec:BCG}
Several groups have considered the degree to which cluster galaxies
alter the cross section for producing giant arcs.  \citet{Mene00} and
\citet{Flores00} studied the influence of cluster galaxies (besides
the BCG), and found them to be generally unimportant.  The effects of
the central brightest cluster galaxy (BCG) were considered by
\citet{Mene03b}, DHH, and \citet{Shirley04}. These 
studies found that BCGs increase the lensing cross section 
for arc separations small enough that the mass enclosed within the 
arc radii has a significant baryonic component. Specifically, 
DHH and \citet{Shirley04} found that for arc radii $\gtrsim
10^{\prime\prime}$, the difference in the number of arcs for pure dark
matter clusters versus clusters including BCGs was not dramatic.

In order to account for the effects of BCGs on the giant arc cross
sections, we artificially add baryons to the centers of each
cluster by `painting' BCGs onto the dark matter surface
density. Similar to the procedure in DHH, we model the mass profile
with singular isothermal spheres (SIS), which accurately represent
strong lensing by elliptical galaxies \citep[see e.g.][]{Koch00}. 
There it was found that the BCG enhancement of the cross section depended
sensitively on the concentration of the central galaxy, parameterized
by the velocity dispersion of the SIS; whereas varying the total mass
of the central galaxy (i.e. the extent of the SIS) had a negligible
effect. We thus choose to keep the mass of the central galaxy to be a
fixed fraction $M_{\rm baryon}=0.003~M_{\rm FOF}$ of the mass of the
dark halo.

In the absence of a theory of BCG formation, we use a simple
prescription for assigning velocity dispersions to the BCGs, scaling
it with the mass of the dark matter halo.  \citet{ES91} found a strong
correlation between the optical luminosity of BCGs and the X-ray
temperature of a large sample of galaxy clusters, with scaling $L_{\rm
  BCG}\propto T_{\rm X}^{8/10}$. Because BCGs lie on the fundamental
plane of elliptical galaxies \citep{OH91}, they obey the \citet{FJ76}
relation $L\propto \sigma^4$. Combining these scaling relations with
the mass--temperature relation for X-ray clusters, $T_{\rm X} \propto
M^{2/3}$, gives $\sigma \propto M^{2/15}$. Normalizing this
relationship to the Coma cluster gives \be \left(\frac{\sigma}{300
  \kms}\right)=\left(\frac{M_{\rm FOF}}{10^{15} \hmsol}\right)^{2/15},
\ee where we took the velocity dispersion of the BCG in Coma to be
$\sigma_{\rm}=323 \kms$ \citep{FIF95} and converted the X-ray
temperature of Coma, $T_{\rm X}=8.0 \kev$ \citep{ES91}, to a FOF group
mass $(b=0.2)$ using the prescription described in
\citet{White01} \citep[see also][]{HK03}.

\section{Quantifying the Effect of Matter in the Light Cone}
\label{sec:light_cone}

\begin{figure}
  \centerline{
    \epsfig{file=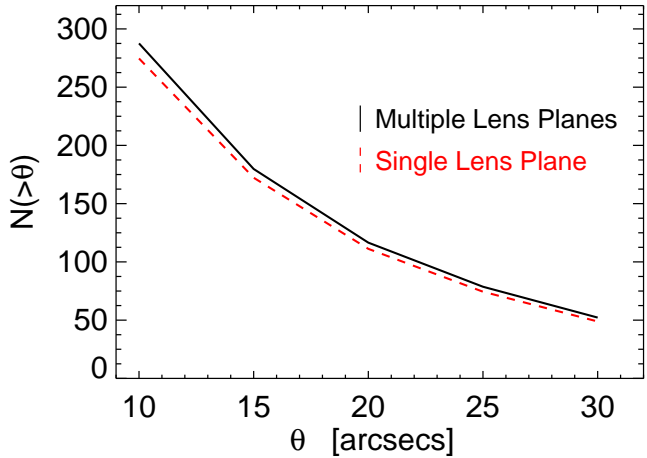,bb=40 10 504 360,width=0.50\textwidth}}
  \caption{ Cumulative distribution of giant arc separations for both
    single plane (dashed red) and multi plane (solid black) ray
    tracing simulations.  Our full ray-tracing simulations show that
    line of sight projections have a modest effect, $\lesssim 7\%$, on
    the statistics of giant arcs with separations $\theta \gtrsim 10\arcsec$.
    \label{fig:light_cone}}
\end{figure}

A question of recent interest has been the importance of projections
of large-scale structure in the line of sight towards strong lenses.
\citet{Wamb04b} have suggested that projections of large scale
structure significantly ($\sim 20-30\%$) increase strong lensing cross
sections, using heuristic arguments.
In this section, we quantify the effect of projections
on giant arc cross sections using full ray tracing simulations. 

We generalized our ray tracing code to compute lensing by matter at
different redshifts using the multiple lens plane algorithm \citep[see
  e.g.][]{sef}. This allows us to include the strong lensing effect of
the large scale structure in the light cone between the source plane
and the observer. This code was used in \citet{DHB05} to study the
effect of projections of large scale structure on strong lensing
cosmography, and a full description can be found there. Here, briefly 
summarize the essential features of the code. 

We sliced through our $320~\hMpc$ simulation cube to produce lens
planes spaced every $160\,h^{-1}$Mpc with angular size roughly
20$\arcmin$, so that the physical side length of the projected planes
telescopes with increasing distance from the observer at $z=0$.  A
total of 243 pairs of planes were produced for each of 19 redshifts
between $z=0-6.37$.  The cosmic mean density was subtracted from each
plane, and the lightcone was tiled by randomly selecting a pair of
planes for each redshift, resulting in a total of 38 lens planes per
realization.  This procedure and the simulation outputs 
are identical to those used by \citet{Wamb04a,Wamb04b}. 

Ideally, we would compute arc cross sections simply by generating many
such realizations and ray tracing over the entire $20\arcmin$ field.
However, because the computation of reliable mean cross sections
requires averaging over a large number of different cluster
orientations, this approach would not be computationally feasible: we
would spend the vast majority of our time ray 
tracing blank regions of sky far from any clusters which would contain
no giant arcs. To circumvent this practical constraint, we simply
insert an extra lens plane containing the massive clusters for the
lens redshift or snapshot that we consider. We ray trace only a
$4~\hMpc$ FOV around each cluster and project through the same number
of different orientations as described in the previous
section. However, each orientation now also includes a different
random realization of the 38 lens planes to account for the large
scale structure along the line of sight.
This allows us to compute the total orientation-averaged
strong lensing cross section for the clusters in this snapshot, but
with the additional effect of large scale structure in the light
cone. A direct comparison can then be made to the ray traces which
included the clusters alone, as the orientation averaging is 
identical for both cases. We will refer to the ray traces which include
large scale structure along the line of sight as the `Multi-Plane'
simulation and those which include only the clusters as the
`Single-Plane'. An example of the critical curves of a massive cluster
with and without the line of sight lens planes is shown in Figure~3 of
\citet{DHB05}. There it can be seen that the critical curves are only
slightly affected by the line of sight density fluctuations.
Note that in order to accurately compute the angular separations of
the arcs from the cluster, we must take into account the deflection of
the cluster center, which can be as large as $\sim 1\arcmin$, by lensing
from the large scale structure between the observer and lens redshift.

We ray traced through a total of 13,594 unique projections of the 878
clusters in our simulation volume for the snapshot at $z_{\rm
  d}=0.41$. The source plane redshift was set to $z_{\rm s}=2.0$. BCGs
were not added to the dark matter mass distribution for this analysis.
In Figure~\ref{fig:light_cone} we compare the cumulative distribution
of giant arc separations for the `Multi-Plane' and `Single-Plane' ray
tracing simulations. It is clear that large scale structure
projections do not have a significant effect on the abundance of giant
arcs, as the largest difference between the two abundances is
$\lesssim 7\%$.  \citet{Wamb04b} reported a larger effect, $\sim
25\%$, for a similar source redshift plane, based on qualitative
arguments regarding the fraction of surface mass density pixels which
became super-critical because of contributions from multiple lens
planes.  Their larger frequency of multi-plane lensing are most likely
smaller separation, $\theta \lesssim 10\arcsec$, galaxy scale lensing
events which would not be considered giant arcs.  Our full ray-tracing
simulations show quantitatively that line of sight projections have a modest
effect on the statistics of giant arcs with
separations $\theta \gtrsim 10\arcsec$.

\section{Analog Halos}
\label{sec:analog}

To elucidate which properties of CDM halos are most important for
strong lensing, we have measured structural properties for our
simulated clusters and generated synthetic analog halos which preserve
one or more of these properties. By comparing the lensing strengths of
the synthetic analog halos to the N-body halos, we are able to
quantify the significance of various aspects of halo structure for
strong lensing.  

The key structural characteristics we measure for our halos are the
radial density profile $\rho(r)$ measured with respect to the most
bound particle, the inertia tensor of the particle distribution, and
the degree of substructure.  We have generated four types of analog
halos: 
\begin{enumerate}
\item {\bf Spherical} analogs preserve the radial density profile $\rho(r)$
  but are spherically symmetric.
\item {\bf Triaxial} analogs are similar to spherical analogs, however
  they preserve the inertia tensor of the cluster core.  
\item {\bf Clumpy core} analogs are similar to triaxial analogs,
  however the particle distribution within the cluster core is left
  unchanged while the mass exterior to the core is smoothed triaxially.
\item {\bf No substructure} analogs preserve the particle
  distribution, except for particles within massive subhalos, which
  are triaxially redistributed.
\end{enumerate}

For the simplest cases of spherical and triaxial analogs which we
describe below, these halos mirror the types of simplified analytical
models which have been used to describe strong lensing by galaxy
clusters previously in the literature \citep{Bart96,%
Mene03a,Oguri03,OK04}. It is well known that these simple models
underpredict the number of giant arcs \citep{Mene03a}. Our analog
halos have the added advantage that they exactly reproduce the joint
distribution of halo shapes and concentrations in the simulation
volume. Thus, they represent the case of the more sophisticated
analytical models which attempt to account for the joint distributions
of halo shapes and concentration measured from numerical simulations
\citep{JS02,Oguri03,OK04}. Although analytical models for the
distribution of substructure in dark matter halos exist
\citep[e.g.][]{Sheth03,SJ03,OL04}, computing strong lensing cross
sections for such models is sufficiently cumbersome that it has not
yet been undertaken. Our ray tracing computation for the 
no-substructure halos allow us to quantify the effect of dense
substructure on the efficiency for producing giant arcs.  Because the
clumpy core analog halos are roughly identical to their parents
inside of an average overdensity of $\sim 10,000$, but are smooth
ellipsoids outside of this contour, a comparison with their parent halos
can quantify the degree to which the structure of the large scale mass
distribution changes strong lensing cross sections. We note that some
of these issues have already been explored in \citet{Mene03a}.  There the
differences between elliptical (triaxial) and real dark matter halos
was quantified via a multipole analysis in Fourier space. Our method
of Analog Halos is complementary, and because we modify these halos
in real space, we can pinpoint the particular halo properties which 
are important to strong lensing. 

\label{sec:analytical}

\begin{figure*}
  \centerline{
    \epsfig{file=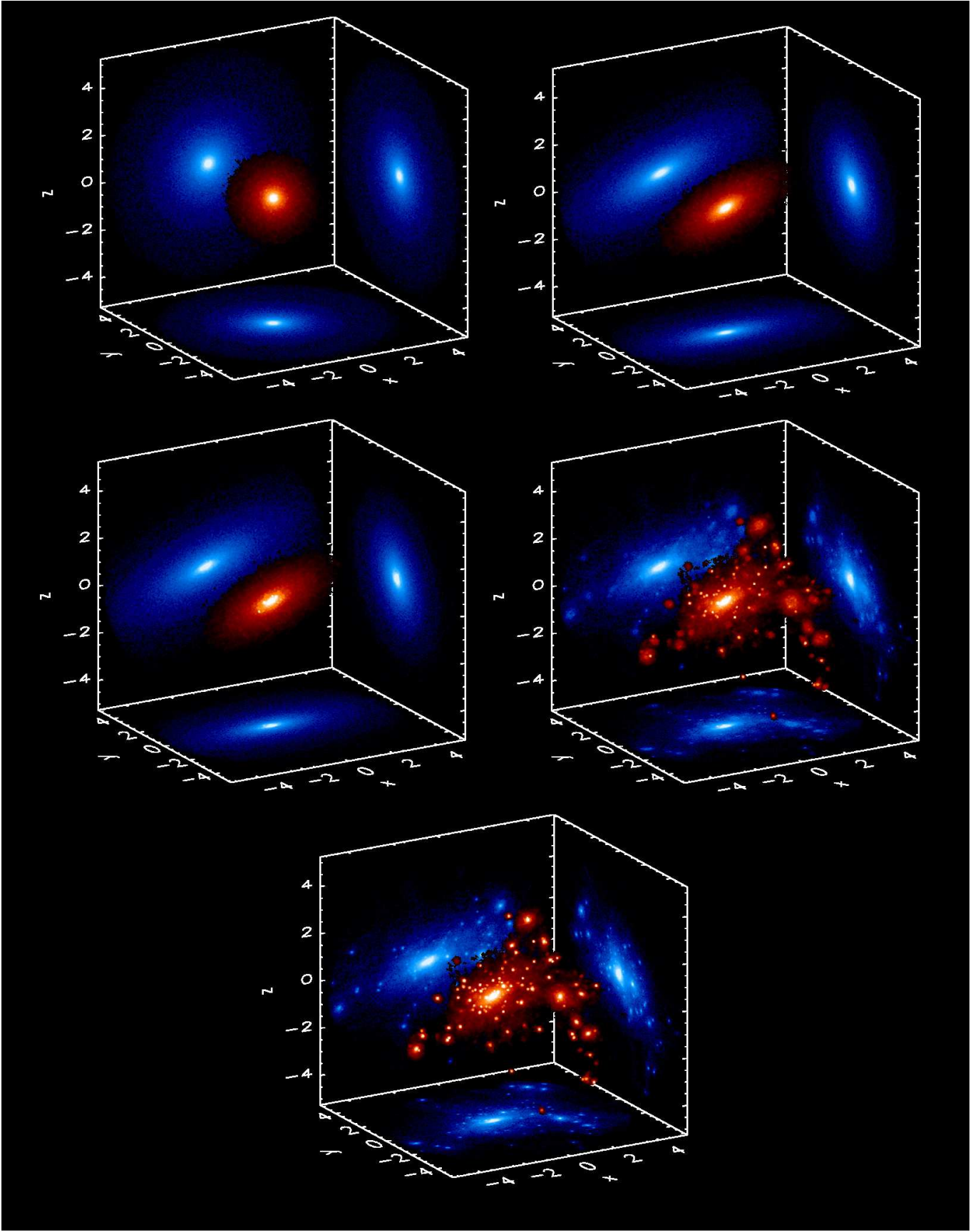,width=1.0\textwidth}}
  \caption{Density distributions of four different types of analog
    halos compared to the original parent.  The surface density for
    three projections are shown on the coordinate axis planes to aid
    visualization. The spatial axes are in units of comoving $\hMpc$
    and both density (2-d and 3-d) colormaps are logarithmically
    stretched.  Upper left is the the spherical analog, the upper
    right is the triaxial analog, middle left is the clumpy core analog,
    middle right is the no substructure analog, and the parent
    halo is depicted at the bottom. The radial mass profiles of the
    analog halos are compared to that of the parent halo in
    Figure~\ref{fig:demo_rho}. 
    \label{fig:demo}
    }
\end{figure*}

\begin{figure}
  \centerline{
    \epsfig{file=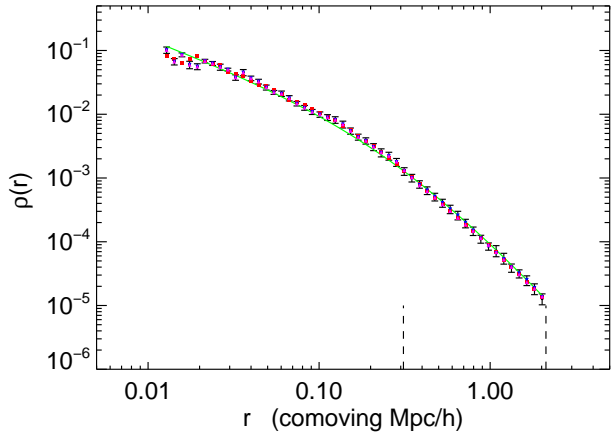,width=0.50\textwidth}}
  \caption{ Comparison of radial density profile $\rho(r)$ of analog
    halos shown in Figure~\ref{fig:demo} to original parent halo. The
    y-axis has units of $\hmsol/\rm {pc}^3$. The black points and
    error bars are the density profile and bootstrapped variances of
    the original parent halo. The solid green line is the best fit NFW
    profile.  The inner and outer dashed vertical lines indicate the
    locations of the scale radius and virial radius for the best fit
    NFW profile, respectively. Red squares, magenta triangles, and
    blue upside down triangles, show the density profile of triaxial,
    and clumpy core analog halos, respectively. The spherical case is
    not shown because its radial density profile is identical to the
    parent halo by construction. Although we kept the ellipsoidal
    radius $R_{e}$ fixed when redistributing particles for the analog
    halos shown, the radial profiles are nevertheless preserved to a
    high degree of accuracy.\label{fig:demo_rho} }
\end{figure}

\begin{figure}[t!]
  \centerline{
    \epsfig{file=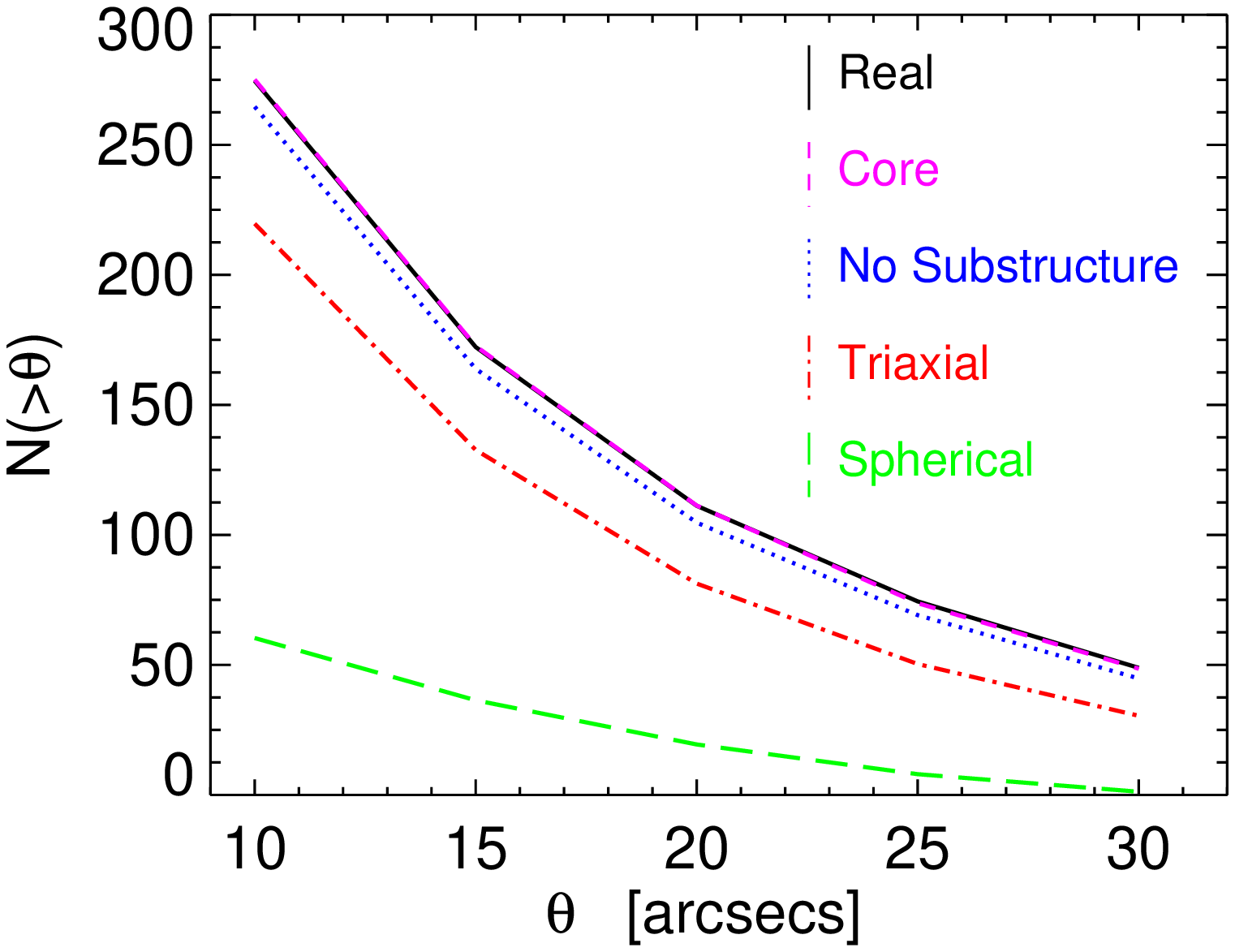,bb=40 0 504 381,width=0.50\textwidth}}
  \caption{ Comparison of the number of giant arcs on the entire sky
    for original and analog halos. The black curve is the cumulative
    distributions of giant arcs lensed by the original 'parent' in the
    snapshot at redshift $z_{\rm d}=0.41$.  The source plane 
    was at $z_{\rm s}=2.0$, and the density of background galaxies in
    the redshift range $[1.75,2.5]$ was computed according to the
    discussion in \S~\ref{sec:gals}. The green (long dashed) curve is
    the distribution of arc separations if all clusters in the
    simulation volume are replaced by their spherical analog
    halos. The red (dot-dashed), blue (dotted) and magenta (short
    dashed) curves show the same quantity but for triaxial, no
    substructure, and clumpy core analog halos, respectively. The data
    plotted above are also shown in
    Table~\ref{tab:analog}. \label{fig:analog}}
\end{figure}

\subsection{Generating Analog Halos}

Spherical analog halos are generated by randomly redistributing the
angular coordinates of particles while holding fixed their radial
coordinates.  The radial profile $\rho(r)$ of each spherical analog is
thus identical to its parent halo. Note that all radii are measured
with respect to the most bound particle in the cluster.

To generate triaxial analog halos we follow a similar procedure as for 
the spherical case. For each particle we compute the ellipsoidal 
radius \citep[see e.g.][]{JS02}
\be
R_e=\sqrt{\frac{X^2}{A^2} + \frac{Y^2}{B^2} + \frac{Z^2}{C^2}},
\label{eqn:Re}
\ee 
as well as $A=(q_2q_3)^{-1/3}$, $B=q_2 A$, and $C=q_3 A$, where the
coordinates $X$, $Y$, $Z$ are measured along the principal axes of the
core (i.e.\ the eigenvectors of the inertia tensor {\bf Q}, see
eqn.~\ref{eqn:Q}).  Each
particle is then randomly mapped onto the surface of the ellipsoid
defined by constant $R_e$. This procedure preserves the density profile in
ellipsoidal coordinates $\rho(R_e)$ \citep{JS02} and also maintains
the axis ratios and principal axes of the core of the original N-body halo.

The no-substructure analog halos are generated by redistributing all
of the particles linked into FOF subhalos ($b=0.05$) with masses
$M_{\rm FOF}>10^{11.5}~\hmsol$, except for those in the largest
subhalo, which are left
untouched.  The subhalo particles are redistributed along the
ellipsoid defined by $R_e$ in eqn.~(\ref{eqn:Re}) using the 
principal axis directions and axis ratios of the core--- i.e. using
exactly the same procedure as for the triaxial analog halos--- thus
also preserving $\rho(R_e)$.  As discussed in \S \ref{sec:sub}, the
FOF algorithm with linking length $b=0.05$ links particles into groups
enclosed by an average overdensity of $\sim 10,000$.  These
analog halos are thus identical to their corresponding N-body halos
except that all substructures containing $M>10^{11.5}~\hmsol$
within a mean overdensity of $~10,000$ have been smoothed out.

We generate the clumpy core analog halos in exactly the same way as
the triaxial case except that the particles in the most massive
subhalo identified by the FOF with $b=0.05$, i.e.\ the cluster core,
are left untouched. Thus the Clumpy Core analog halos are roughly
identical to their parents inside of a mean overdensity of $\sim
10,000$, but smooth and nearly identical to the `Triaxial' analog halos
outside of this overdensity.

Figure~\ref{fig:demo} compares the cluster lens in our simulation with
the largest lensing cross section to all of its different analog
halos.  This cluster, taken from a snapshot at $z_{\rm d}=0.41$, has
mass $M_{\rm vir}=8.1\times 10^{14}~\hmsol$ and concentration $c_{\rm
  NFW}=6.9$.  The axis ratios of the core are $q_2= 0.46$ and
$q_3=0.40$. In Figure~\ref{fig:demo_rho}, we compare the radial
density profile of this cluster to the profiles of the different
analog halos. The spherical case is not shown because its radial
density profile is identical to the parent by construction. Note
that although we keep the ellipsoidal radius $R_{e}$ fixed when
redistributing particles to create the triaxial, no substructure,
and clumpy core analog halos, in practice the radial profiles
$\rho(r)$ are nevertheless preserved to a high degree of accuracy.

\subsection{Lensing Comparison}

We generated each type of analog halo for all 878 clusters in the the
snapshot at $z_d=0.41$. We ray traced through every cluster for a
source plane at $z_{\rm s}=2.0$, averaging over the same orientations
for each halo (see \S~\ref{sec:ray_trace}), and computed the
cumulative distribution of image splittings following
eqn.~(\ref{eqn:Nj}). Since our aim is to understand which characteristics
of the dark matter distribution are most important for strong lensing,
BCGs were not added to the mass distribution for this analysis. In
Figure ~\ref{fig:analog} we compare the number of giant arcs produced
by the real and analog halos, assuming an area equal to the entire
sky.  The data used to construct this figure is given in
Table~\ref{tab:analog}.

From the data in the table, we see that spherical halos underpredict
the abundance of giant arcs by a factor as large as 50 and triaxial
models fall short by a factor as much as $60\%$. These results are
consistent with the findings of \citet{Mene03a}.  Because the
distribution of halo concentrations and shapes of the parent halos is
exactly reproduced in our analog halos, we conclude that even the more
complicated analytical models which convolve analytical cross sections
with the distributions of halo concentrations and shapes
\citep{Oguri03,OK04}, will underpredict the number of cluster lenses
by a large amount.  It is noteworthy that triaxiality increases the
number of giant arcs by a factor of 4-25 compared to the spherically
symmetric halos.  This is because the shallow density cusps $\rho
\propto r^{-1}$ of CDM halos result in an extreme sensitivity to
triaxiality, as has been emphasized by several authors
\citep[DHH;][]{DK03,BM04}.

From a comparison of the real halos to the no substructure analogs, it
is apparent that projections of dense substructure onto the small
radii probed by strong lensing has a relatively modest effect on the
lensing cross sections, increasing the total number of arcs by $\sim
5-10\%$. \citet{Mene00} came to similar conclusions about changes in
lensing cross sections caused by baryonic substructure (cluster
galaxies) which have small masses $\lesssim 10^{12}~\hmsol$.  Here we
find a similar result for dark matter substructure, which can be much
more massive for the extreme case of a binary or merging cluster.  We
revisit the effect of substructure on strong lensing in
\S~\ref{sec:histo}, when we statistically quantify the properties
of cluster lenses.

\begin{table*}[t!]
  \begin{center}
    \caption{Analog Halo Comparison\label{tab:analog}}
    \begin{tabular}{lccccc}
      \hline
      \hline
      Type & $N(\theta >10\arcsec)$ \ & \ $N(\theta >15\arcsec)$ \ & \ $N(\theta >20\arcsec)$ \ &  \ $N(\theta >25\arcsec)$ \ & \ $N(\theta >30\arcsec)$  \ \\
      \hline
      Real      \hfill\vline &   274.6    &  172.3     &  111.2   &   74.4    &     48.9\\
      Clumpy Core      \hfill\vline &   275.1    &  172.6     &  111.3   &   73.8    &     48.5\\
      No Substructure   \hfill\vline &   264.7    &  163.9     &  104.7   &   69.1    &     44.8\\
      Triaxial  \hfill\vline &   219.7    &  132.7     & \phn81.3 &   50.3    &     30.4\\  
      Spherical \hfill\vline &   \phn60.3 &  \phn36.4  & \phn19.4 & \phn8.0  &   \phn1.2\\
    \hline
    \end{tabular}
  \end{center}
  \footnotesize NOTES--- Comparison of the number of giant arcs produced by real halos to the number 
  produced by analog halos. The quantity $N(\theta >10\arcsec)$ is the cumulative number of 
  of giant arcs on the entire sky for a source plane at $z_{\rm s}=2.0$. The data in this table 
  are plotted in Figure~\ref{fig:analog}.
\end{table*}

Recall that the clumpy core analog halos are identical to the original
parent halos inside a mean overdensity of $10,000$, but are smooth
ellipsoids outside of this contour.  The close agreement between the
clumpy core analogs and the parent halos, indicates that halo
substructure and the large scale mass distribution do not
significantly influence strong lensing cross sections. The number of
giant arcs produced by CDM halos is \emph{primarily determined by the
  mass distribution within a mean overdensity of} $\sim 10,000$.

The difference between the cross sections for triaxial halos and the
full N-body halos indicates that some systematic departure from
ellipsoidal symmetry, on average, enhances lensing strength.  Some
possibilities for this include substructure on very small scales, or
perhaps lopsidedness in the form of boxiness or diskiness in the
isodensity contours.  Naively, we would not expect massive
substructures to persist in the high-density cores of clusters since
the dynamical friction timescale is much shorter than the Hubble time.
We have attempted to detect whether larger-scale departures from
ellipsoidal symmetry enhance lensing cross section by running a
principal component analysis on the halo density profiles.  The
results of this analysis were unclear: the principal components we
recovered with high signal to noise appeared to correspond to varying
concentration and ellipticity, with no other significant components
apparent.  This point warrants future study, since it suggests that
the abundance of giant arcs is sensitive to the detailed morphology of
dark matter halos on small scales.

\section{Characterizing The Cluster Lens Population}
\label{sec:histo}

\begin{figure}
  \centerline{
    \epsfig{file=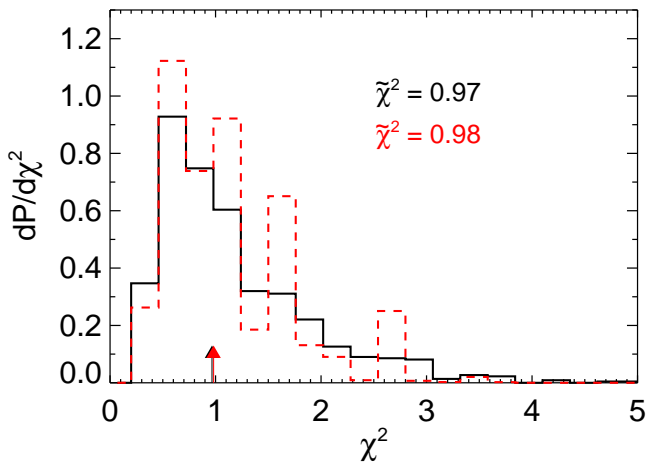,bb=40 0 504 360,width=0.50\textwidth}}
  \caption{ Lensing (dashed) and total (solid) distributions of the
    reduced $\chi^2$ of the NFW fit to the number profile of each
    halo. The median of both the lensing and total distributions,
    indicated by the arrows, are labeled in the upper right . The fact
    that the median value of the total distribution is near unity
    gives us confidence that the NFW profile is providing a good fit
    to the simulated clusters and the bootstrapped $\sigma_i$ are
    accurately describing the errors (see \S~\ref{sec:NFW}). The
    similarity between the lensing and total distribution indicates
    that NFW profiles provide just as good a fit to lensing clusters
    as they do to the total population. \label{fig:chi2_hist}}
\end{figure}

\begin{figure*}
  \vskip 1.2cm \centerline{ \epsfig{file=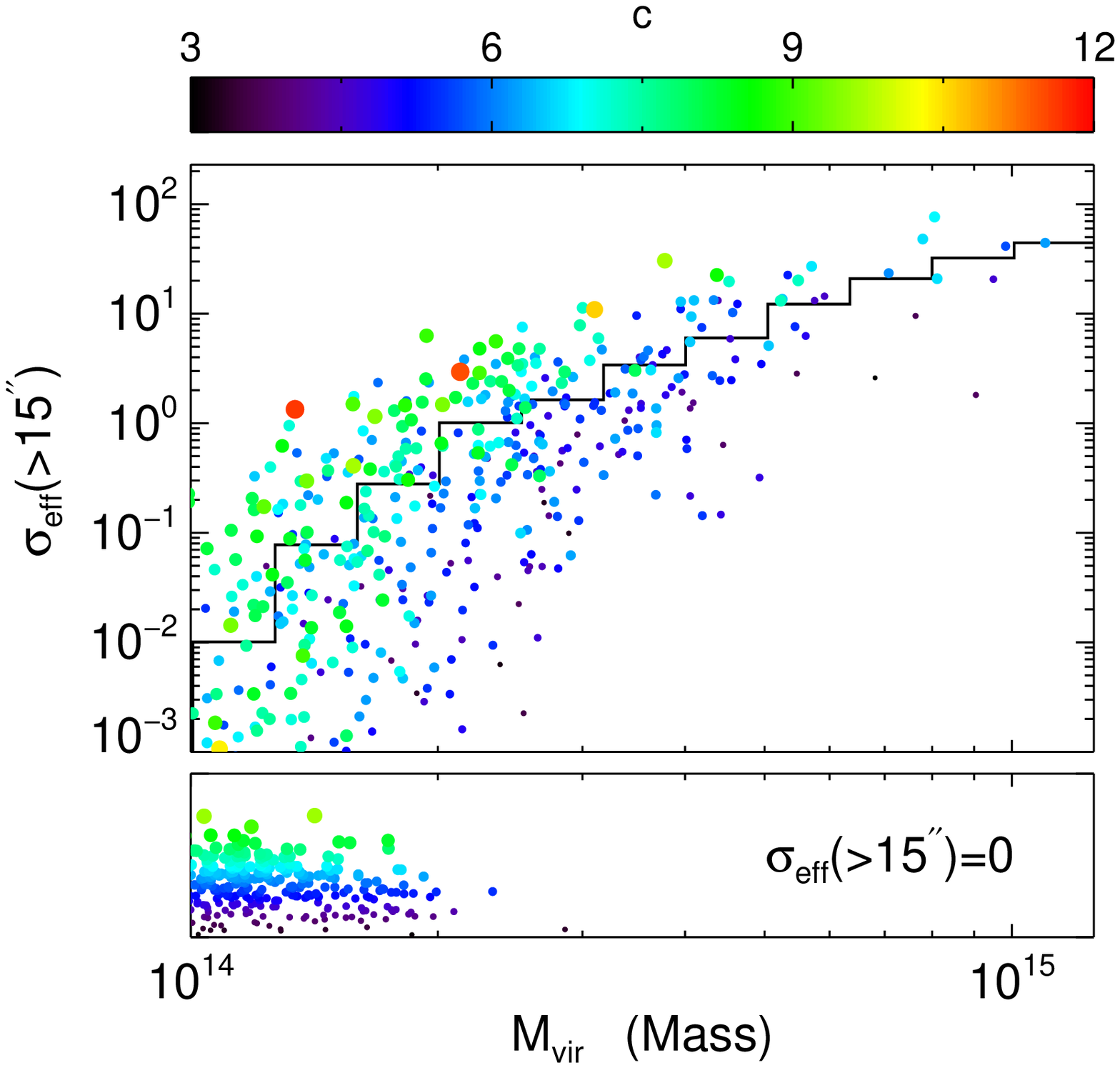,bb=10 10
      500 400,width=0.5\textwidth} \epsfig{file=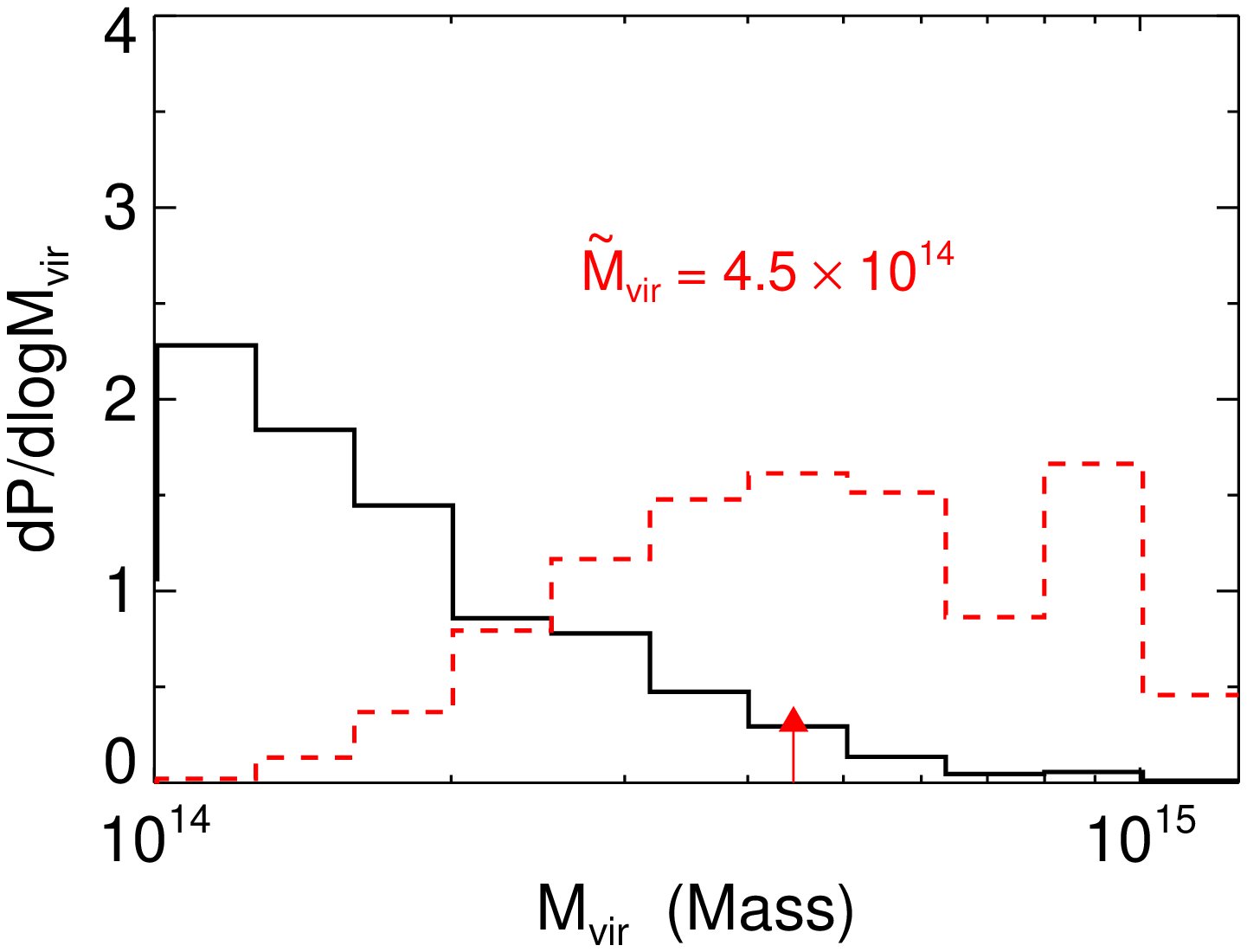,bb=50 15
      504 360,width=0.50\textwidth}}
  \caption{\emph{Left:} Scatter plot of effective strong lensing cross
    section for forming giant arcs with $\theta>15\arcsec$ against
    cluster mass, for clusters at $z_{\rm d}=0.41$.  Each point
    corresponds to a cluster in our simulation volume and the cross
    sections have been averaged over all orientations (see
    \S~\ref{sec:ray_trace}). The y-axis has units of arcsec$^2$, and 
    the colors and sizes of the points indicate
    the value of the lens strength parameters $\kappa_{\rm s}$ (see
    \S~\ref{sec:NFW}). For $M\gtrsim 10^{15}~\hmsol$,
    all clusters are effective strong lenses; whereas at lower
    masses $M\lesssim 3\times 10^{14}~\hmsol$ only the most concentrated
    members of the population can multiply image quasars.  The histogram
    shows the mean strong lensing probability averaged over mass bins of
    width $0.1$ in $\log_{10}M$. Clusters which have zero strong lensing
    cross section for all orientations are plotted in the lower panel,
    where we have stretched the y-axis so that the points do not all
    overlap.  \emph{Right:} Comparison of the probability distribution
    of the virial mass, $P_{\rm lens}(\log_{10}M_{\rm vir})$ of cluster
    lenses (dashed red curve) to the mass function, normalized to unity
    over the mass range considered.  (solid black curve). These are the
    same clusters at $z_{\rm d}=0.41$ shown in the left panel. 
    The bin spacing is the same as in the left panel. The red
    histogram $P_{\rm lens}(\log_{10}M)$ can be obtained from the
    scatter plot in the left panel by summing the values of the cross
    sections in each mass bin and dividing by the total cross section
    integrated over all bins (see eqn.~\ref{eqn:Pu}). The mean value of
    $\log_{10}M$ of the cluster lens sample is $M_{\rm vir}=10^{\langle
      \log_{10} M\rangle_{\rm lens}}=4.5\times 10^{14}~\hmsol$,
    indicated by the arrow.  \label{fig:scatter}}
\end{figure*}

In this section we address the following question: How different is
the population of lensing selected clusters from the general cluster
population? Because cluster strong lensing is sensitive to the mass
distribution interior to the NFW scale radius $r \lesssim 300~\hkpc$,
it provides a means to measure the concentration of clusters,
especially when combined with larger scale mass measurements like weak
lensing \citep{Kneib03,Gavazzi03,Broad05b}.  It is thus interesting to
quantify the degree to which these concentrations will be
systematically higher than the typical cluster in the
Universe. Besides concentration, other biases might exist. Are lensing
clusters significantly more or less triaxial?  Do they have more/less
substructure? How likely is the major axis of a lensing cluster to be
aligned with the line of sight?

We are in a position to answer these statistical questions because of
the large number of clusters and orientations that we have ray traced.
Specifically, for the snapshot at $z_{\rm d}=0.41$ we ray traced
through a total of 13,594 unique projections of the 878 clusters in
our $V=(320~\hMpc)^3$ simulation volume. This constitutes an increase
of two orders of magnitude in both the number of clusters and number
of orientations of any previous study.  For the remainder of this
section, we focus our statistical analysis on the clusters in this
snapshot at $z_{\rm d}=0.41$. Note that BCGs were added to the mass
distributions of these clusters as described in \S~\ref{sec:BCG} and
thus are included in the strong lensing cross sections. However, we
measured the cluster parameters (mass, concentration, triaxiality)
from the dark matter particle distributions alone.

We define the probability distribution of the cluster parameter $u$
(such as mass, concentration, axis ratio, etc.)
\be 
P_{\rm
  lens}(u)=\frac{1}{\sum \sigma_{\rm eff}}\frac{d\sigma_{\rm eff}}{du},
\label{eqn:Pu}
\ee 
where $\sum \sigma_{\rm eff}$ is the total effective cross section
of all the clusters (see eqn.~\ref{eqn:sigeff}) in our simulation
volume and $\frac{d\sigma_{\rm eff}}{du}$ is the differential
distribution of the statistic $u$, which we compute by summing the
total cross section with $u$ between $u$ and $u+\Delta u$.  In what
follows, we use $\sigma_{\rm eff}(\theta > 15\arcsec)$, the effective
cross section for forming giant arcs with separations $>15\arcsec$.  A
comparison of the lensing distribution $P_{\rm lens}(u)$, to the
probability distribution of the total cluster population $P_{\rm
  tot}(u)$ will indicate the degree to which the lenses constitute a
biased sample.

In what follows we will compare the NFW parameters of cluster lenses
to the total cluster population. This comparison is clearly only
sensible provided that NFW profiles provide a good fit to the cluster
lenses. In Figure~\ref{fig:chi2_hist} we compare the distribution of
the reduced $\chi^2$ of the NFW profile fits (see eqn.~\ref{eqn:chi2})
for the cluster lenses to the those for the total population.  The
similarity of the distributions indicates that NFW profiles provide
just as good a fit to lensing clusters as they do to the total
population. Furthermore, the fact the reduced $\chi^2\sim 1$ indicates
that NFW profiles provide good fits to the mass distributions of our
clusters.

\subsection{Mass and Concentration}

In the left panel of Figure~\ref{fig:scatter} we show a scatter plot
of the orientation-averaged effective cross sections 
$\sigma_{\rm eff}(\theta > 15\arcsec)$ against
virial mass, for all 878 of the clusters in the snapshot at $z_{\rm
  d}=0.41$.  The colors and sizes of the points indicate the value of
the concentration of each cluster.  The histogram shows the strong
lensing probability averaged over mass bins of width
$d\log_{10}M=0.1$.  Sub-critical clusters which had zero effective
cross sections are plotted in the lower panel, where we have
arbitrarily stretched the vertical axes so the points do not
overlap. The scatter plot illustrates the range of cross sections
which exist in a given mass bin. In addition, it shows that for the
largest masses characteristic of superclusters, $M\sim 10^{15}~\hmsol$,
all clusters are effective strong lenses, whereas at lower masses
$M\lesssim 3\times 10^{14}~\hmsol$, only the more concentrated
members of the population are strong lenses.  Notice the
extremely steep dependence of the lensing cross section on mass: the
mean cross section changes by nearly four orders of magnitude over a
single decade in mass.  In the right panel, we compare the probability
distribution $P_{\rm lens}(\log_{10}M)$ (dashed red curve) to the
cluster mass function normalized to unity over the mass range
$M>10^{14}~\hmsol$ considered (solid black curve).  The bin spacing in
the right panel is the same as in the left panel. The red histogram
$P_{\rm lens}(\log_{10}M)$ can be obtained from the scatter plot in
the left panel by summing the values of the points in each mass bin
and normalizing the histogram by the total number of lenses (see
eqn.~\ref{eqn:Pu}). The median value of $\log_{10}M$ for the cluster
lens sample is $M=10^{\log_{10}\tilde{M}_{\rm lens}}=4.5\times
10^{14}~\hmsol$.

The left panel of Figure~\ref{fig:scatter} suggests that lensing
selected cluster samples are likely to show significant concentration
bias.  Since we have seen that cross section is such a steep function
of cluster mass, it is desirable to first normalize out the mass
dependence of cluster concentration. Thus we consider the distribution
of the ratio $c_{\rm vir}\slash c_{\rm vir}(M)$, where $ c_{\rm
  vir}(M)$ is some kind of average concentration.  Because simply
computing the mean concentration in mass bins will be sensitive to
outliers and very noisy at the high mass end where we have few
clusters, we must be careful about the fitting procedure. In
Figure~\ref{fig:conc} we show a scatter plot of concentration
versus virial mass measured from the clusters in our simulation volume
at $z_{\rm d}=0.41$. The blue curve is a second order B-spline fit to
points with breakpoints set every 175 clusters ordered by mass. This
fit was computed iteratively with $3\sigma$ outliers rejected at each
iteration until convergence was achieved. \citet{Bullock01} measured
the trend $c=\frac{9}{(1+z)}(M/M_{\ast})^{-0.13}$ from a large
ensemble of numerically simulated clusters, where $M_{\ast}$ is the
nonlinear mass at $z=0$ which is $M_{\ast}=1.3\times 10^{13}~\hmsol$
for our cosmology. The red dashed curve in Figure~\ref{fig:conc}
curve shows the result of adopting the \citet{Bullock01} power law
scaling and fitting the median concentration in five logarithmically
spaced mass bins for the linear amplitude. We get
$c=\frac{12.3}{(1+z)}(M/M_{\ast})^{-0.13}$, i.e. our average
concentrations are $37\%$ higher than \citet{Bullock01}\footnote{This
  disagreement is likely due to differences in the halo fitting
  procedure and also possibly because of the different cosmological
  model simulated. \citet{Bullock01} truncate halo radii when halos
  overlap and they only fit particles \emph{bound} to the halo. We
  argued in \S~\ref{sec:NFW} that neither practice is appropriate to
  the context of gravitational lensing.}. Our B-spline fit is in near
agreement with the relation found by \citet{Bullock01}, although the
mass scaling is not quite as steep.

The lensing and total distributions of $c\slash c(M)$ are shown in the
right panel of Figure~\ref{fig:conc}. The median of the total
distribution is consistent with unity to better than $1\%$, indicating
that the average concentration shown in the left panel of
Figure~\ref{fig:conc} is a good fit to $c(M)$. We see that the cluster
lens population has three dimensional concentrations $\sim 18\%$
higher than the average cluster at a similar mass.

\begin{figure*}
  \centerline{
    \epsfig{file=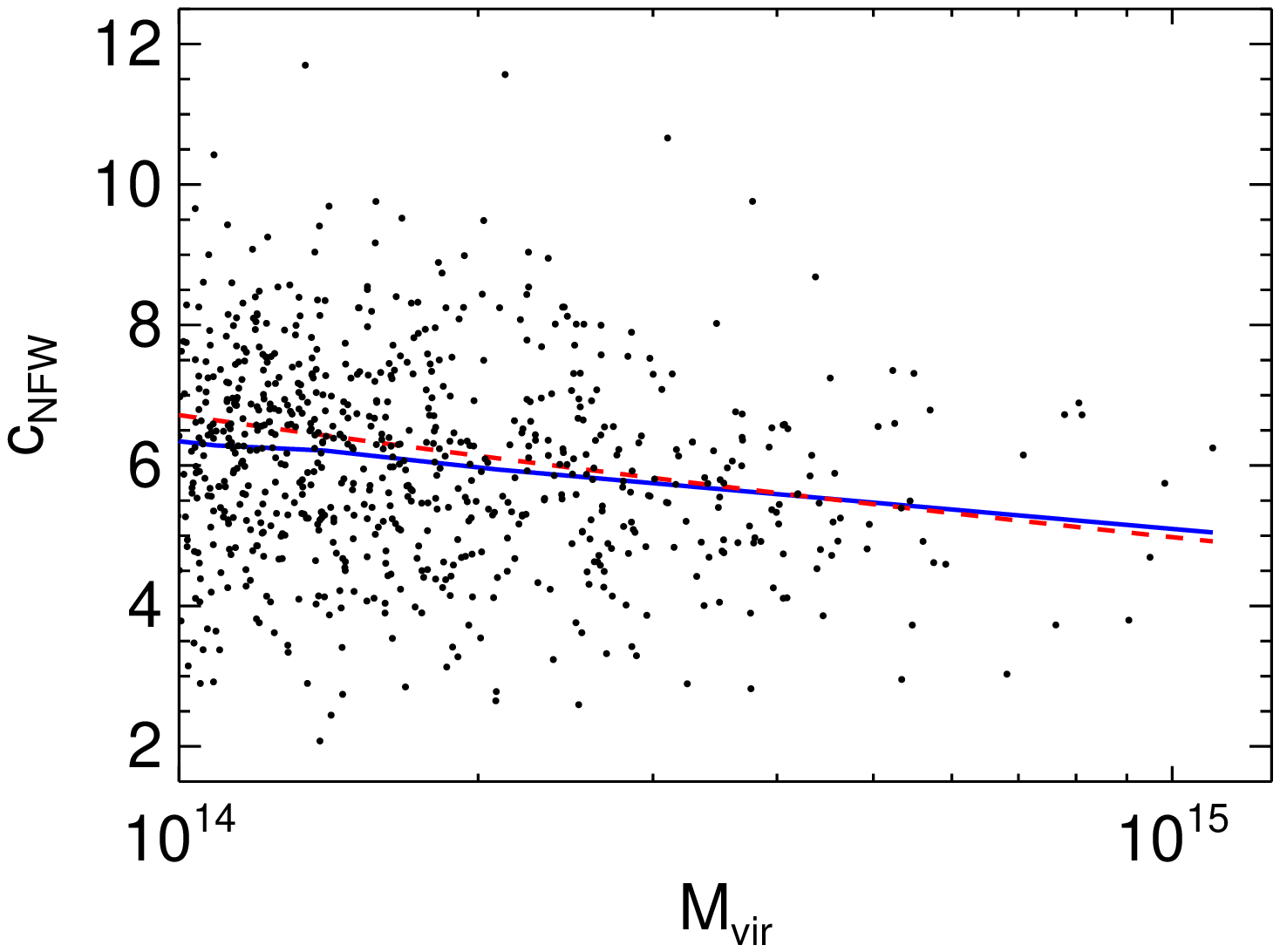,bb=50 0 504 360,width=0.50\textwidth}
    \epsfig{file=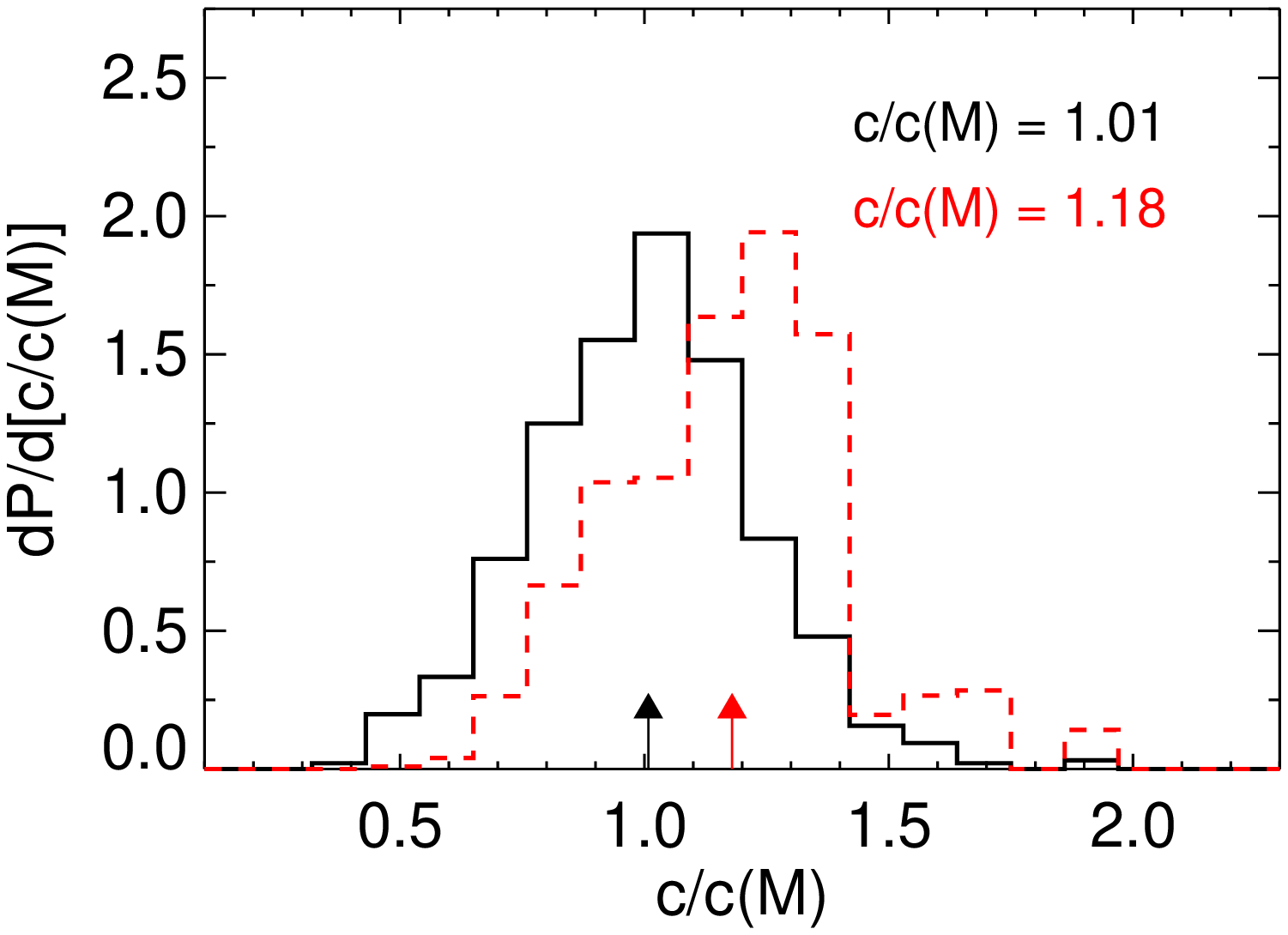,bb=40 0 504 360,width=0.50\textwidth}}
  \caption{ \emph{Left:} Scatter plot of cluster concentration versus
    virial mass for all the clusters in our simulation volume at
    $z_{\rm d}=0.41$. The solid blue curve is a second order B-spline
    fit to the points with breakpoints set every 175 clusters ordered
    by mass. To reduce sensitivity to outliers, this fit was computed
    iteratively with $3\sigma$ outliers rejected at each iteration
    until convergence was achieved. The dashed red curve shows the
    fit, $c_{\rm vir}=\frac{12.3}{(1+z)}(M/M_{\ast})^{-0.13}$, which
    is the result of adopting the \citet{Bullock01} power law scaling
    and fitting the median concentration in five logarithmically
    spaced mass bins for the amplitude. The B-spline fit (solid blue
    line) is in near agreement with the relation found by
    \citet{Bullock01}, although the mass scaling is not quite as
    steep.\label{fig:conc} \emph{Right:} Lensing (dashed) and total
    (solid) distributions of the ratio $c\slash c(M)$. Arrows indicate
    the median of the distributions. The fact that the median of the
    total population is consistent with unity indicates that our fit
    for $c(M)$ has effectively removed the average mass scaling from the
    concentration.  Cluster lens have three dimensional concentrations 
    $18\%$ higher than the typical cluster with similar mass. \label{fig:conc}}
\end{figure*}
\begin{figure}
  \centerline{
    \epsfig{file=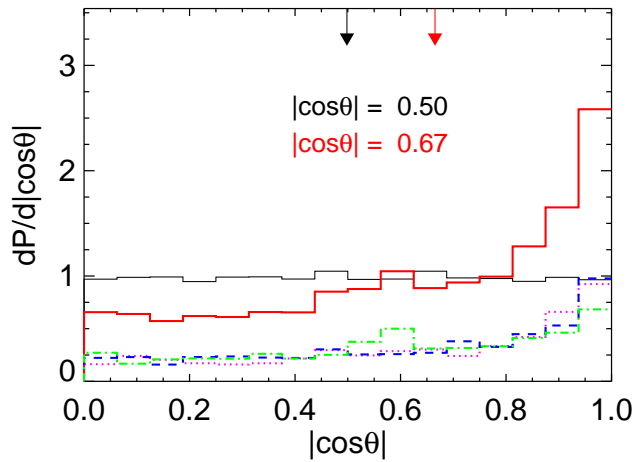,bb=40 0 504 360,width=0.50\textwidth}}
  \caption{ The lensing and total distributions of the absolute value
    of the cosine of the angle between the line of sight and the major
    axis direction of the cluster. The thin (black) line is the
    intrinsic distribution, which is flat as expected. The axis ratio
    $q_2$ is divided into three groups with equal lensing
    probabilities. The magenta (dotted), blue (dashed), and green
    (dot-dashed) curves are individual contributions from
    clusters with axis ratio $q_2$ in the lower third $q_2<0.50$ (most
    triaxial), middle third $0.50<q_2<0.66$, and upper third
    $q_2>0.66$ (least triaxial). These contributions sum to give the
    total distribution which is the thick solid (red) curve.  The median
    of both the lensing and total distributions, indicated by the
    arrows, are labeled on the plot.  The median for the lensing
    clusters is $\left|\cos\theta\right|=0.67$, indicating
    significant alignment bias.
    \label{fig:costheta}
  }
\end{figure}

\subsection{Triaxility and Orientation Bias}

In this section we investigate alignment bias and study the effect of
triaxiality on strong lensing probability. The lensing and total
distributions of the absolute value of the cosine of the angle between
the line of sight and the major axis direction of the cluster,
$\left|\cos\theta\right|={\bf {\hat z}}\cdot{\bf {\hat V}}_{a}$, are
shown in the left panel of Figure~\ref{fig:costheta}. The thin (black)
line is the distribution for all clusters, which is flat as expected.
The axis ratio $q_2$ is divided into three groups with equal total
lensing probability.  The magenta (dotted), blue (dashed), and
green (dot-dashed) curves are individual lensing contributions from
clusters with axis ratio $q_2$ in the lower third $q_2<0.50$ (most
triaxial), middle third $0.50<q_2<0.66$, and upper third $q_2>0.66$
(least triaxial). These contributions sum to give the total lensing
distribution, which is the thick red (solid) curve. The median value
for the lensing clusters is $|\cos\theta|_{\rm lens}=0.67$,
indicating significant alignment bias. Furthermore, the
alignment bias is larger for the more triaxial clusters, as expected.
Our findings
disagree with the discussion in \citet{BSW95} who also compared the
orientation of the cluster principal axes to the line of sight and
found no significant correlation between the two.  Because this study
only ray traced 3 different projections through 13 clusters, it is
likely that they did not have the statistics to measure the
correlation which we detect with high significance.

\begin{figure*}
  \centerline{
    \epsfig{file=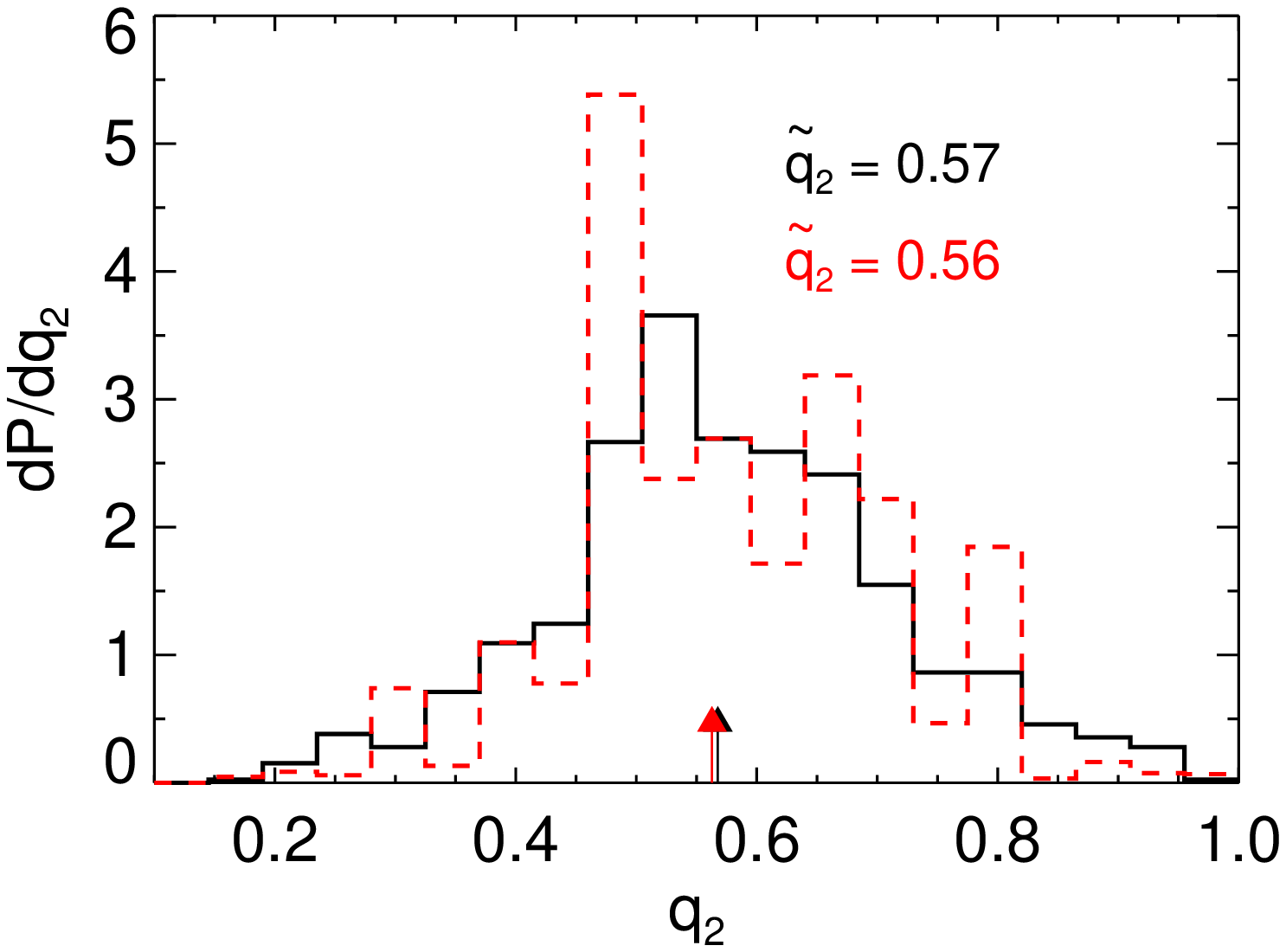,bb=40 0 504 360,width=0.50\textwidth}
    \epsfig{file=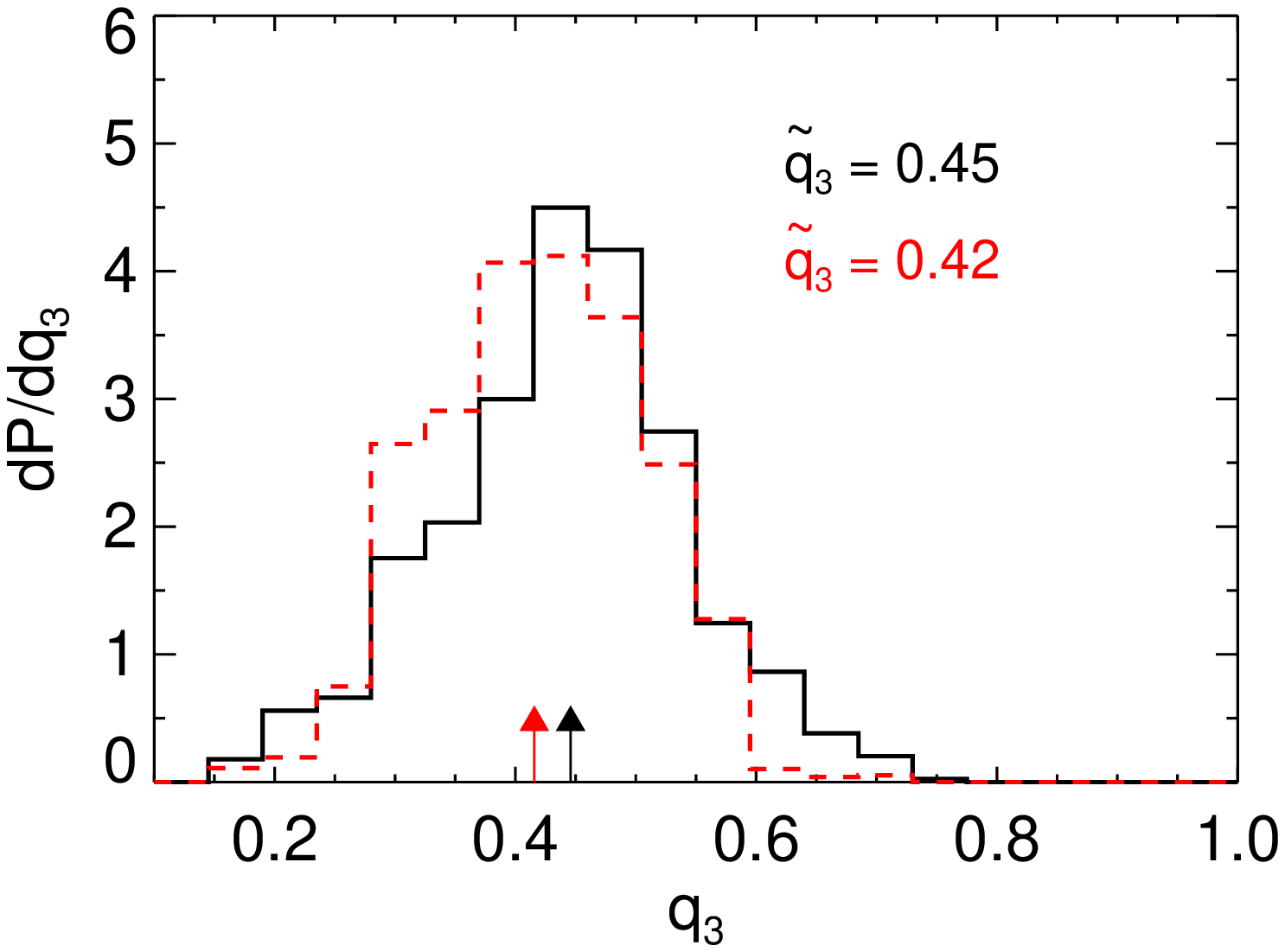,bb=40 0 504 360,width=0.50\textwidth}}
  \caption{ Lensing (dashed) and total (solid) distributions of the
    axis ratios. The left panel shows the distributions of the axis
    ratio $q_2$ and the right panel shows the distributions of $q_3$.
    The mean of both the lensing and total distributions, indicated by
    the arrows, are labeled in the upper right corner of each
    plot. The lensing population has nearly the same distribution of
    axis ratios as the total cluster population.
    \label{fig:q_dist}
  }
\end{figure*}

Despite this large tendency for lensing clusters to be aligned with
the line of sight, the lensing population has nearly the same
distribution of axis ratios as the total cluster population. This is
the clear indication of Figure~\ref{fig:q_dist}, which shows the
distributions of the axis ratios $q_2$ and $q_3$ (shown in the left
and right panels, respectively). The lensing distributions are
nearly indistinguishable from the total cluster population.

We have detected significant alignment bias between
the line of sight and the principal axis direction of the clusters, and
furthermore we argued in \S \ref{sec:analog} that triaxiality significantly 
increased the total lensing cross section for CDM clusters.   On the 
other hand, Figure~\ref{fig:q_dist} indicates that lensing
clusters have the same distribution of axis ratios as the total cluster
population. These two statements seem to contradict each other. If triaxiality
enhances lensing cross sections and if their is a tendency for strong 
lenses to be aligned with the line of sight, why are lensing 
clusters not systematically more triaxial than the total population?

\begin{figure}
  \centerline{
    \epsfig{file=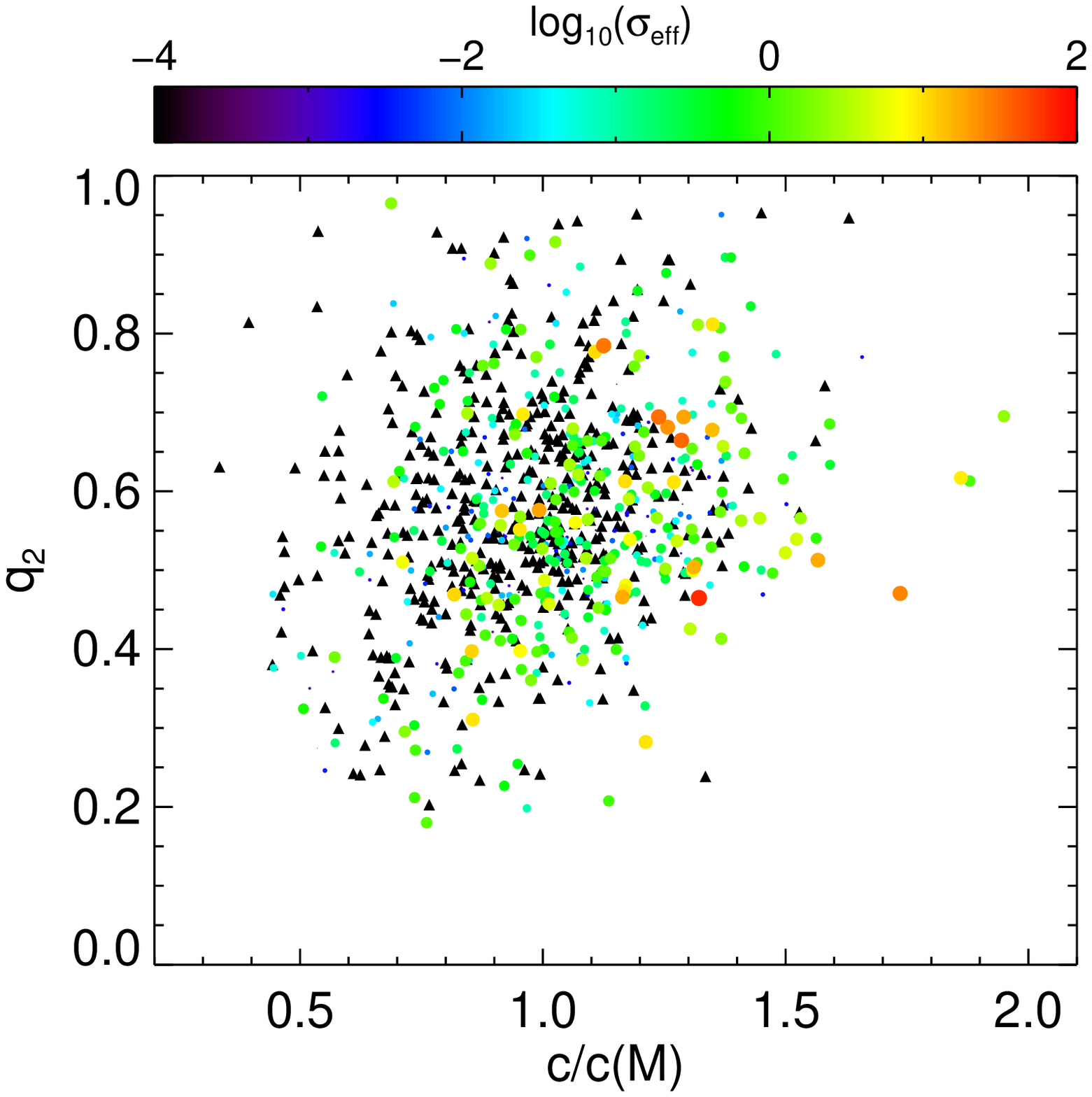,bb=50 0 504 468,width=0.50\textwidth}}
  \caption{ Scatter plot of normalized concentration, $c\slash c(M)$
    versus axis ratio $q_2$ for the clusters in our simulation at
    $z_{\rm d}=0.41$. The black triangles show clusters with zero
    strong lensing cross section $\sigma_{\rm eff}(>15\arcsec)$ and
    circles show clusters with nonzero cross sections, where the
    colors and sizes of the points indicate the size of the cross
    section in units of arcsec$^2$. Cluster with a large degree of
    triaxiality (small $q_2$) tend to be under concentrated; whereas,
    clusters which are less triaxial (large $q_2$) tend to be over
    concentrated. This correlation between $q_2$ and normalized
    concentration is expected on dynamical grounds (see discussion in
    \S~\ref{sec:histo}).
    \label{fig:cvsq}}
\end{figure}

One possible explanation for this apparent contradiction is that the
shape of a dark halo is correlated with its concentration, as might be
expected on dynamical grounds. \citet{Allgood05} have found that
simulated dark halos which have recently collapsed are more triaxial
($q_2$ smaller) and less concentrated, whereas halos which have
collapsed in the more distant past are on average less triaxial ($q_2$
larger) and more concentrated. Dynamically old clusters formed when
the Universe was more dense and hence are more concentrated, and
similarly have had more time to relax and hence should be less
triaxial.  Indeed, a similar trend exists in our sample of clusters as
is shown in the scatter plot of normalized concentration versus axis
ratio in Figure~\ref{fig:cvsq}.

\subsection{2-D vs.\ 3-D concentration}
\label{sec:conc2d}

\begin{figure*}
  \centerline{
    \epsfig{file=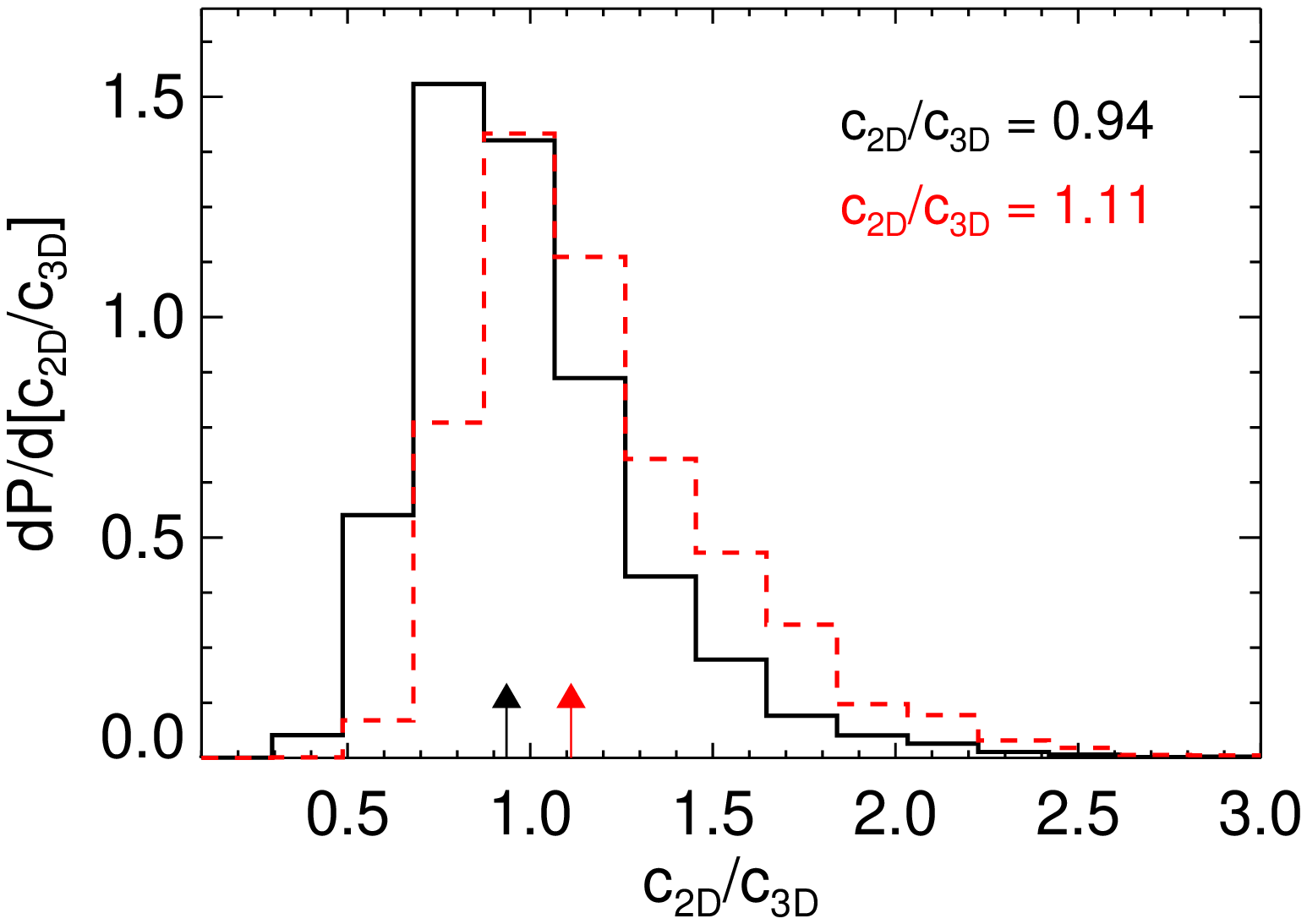,bb=50 0 504 360,width=0.50\textwidth}
    \epsfig{file=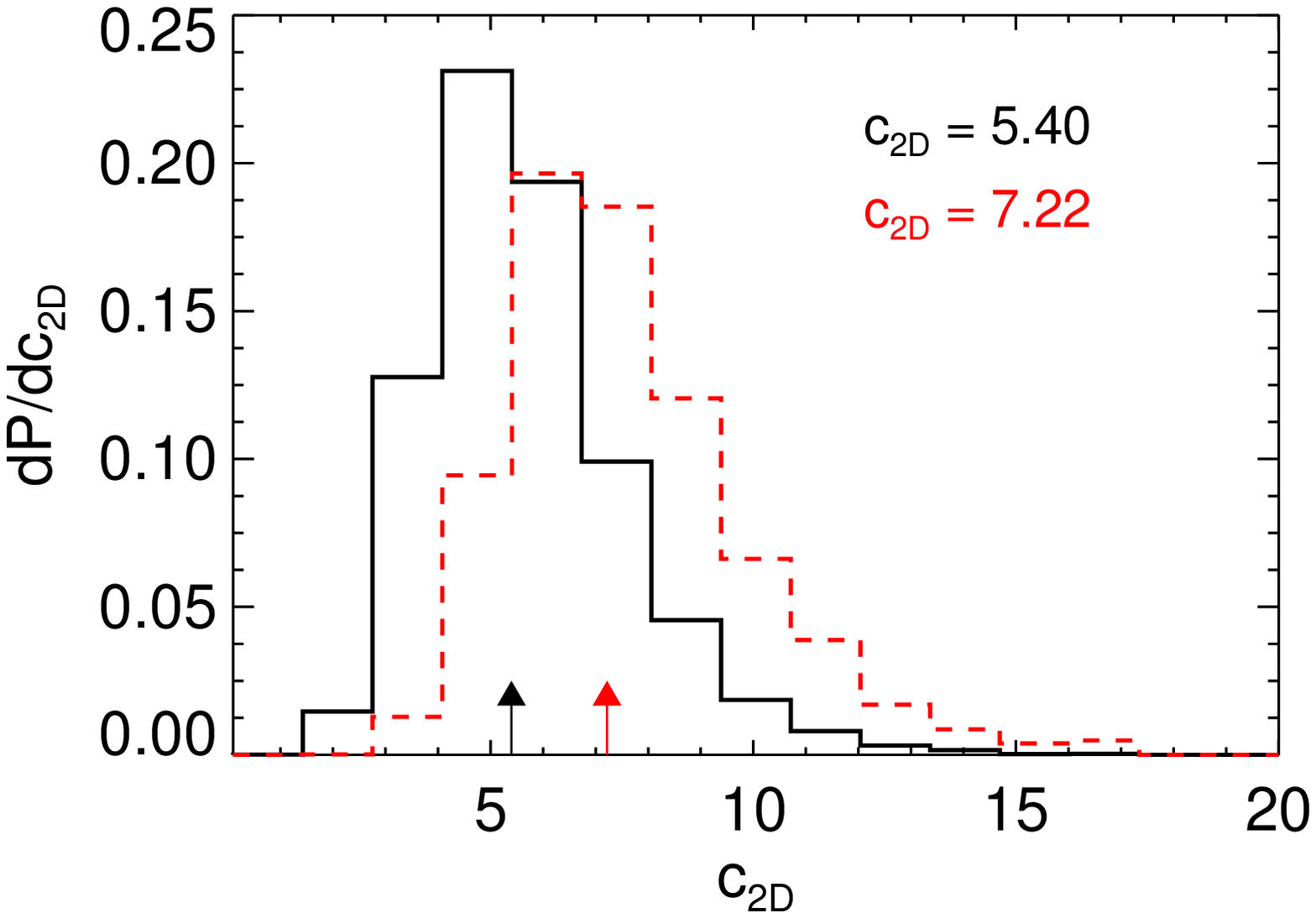,bb=40 0 504 360,width=0.50\textwidth}}
  \caption{ \emph{Left:} Lensing (dashed) and total (solid)
    distributions of the ratio $c_{\rm 2D}\slash c_{\rm 3D}$ where the
    two dimensional concentrations have been measured for each
    projection through the clusters. Arrows indicate the median of
    each distribution. Strong lensing prefers the projections through
    clusters which give $\sim 19$\% higher concentrations.
    \emph{Right:} Lensing (dashed) and total (solid) distributions of
    the two dimensional concentrations, $c_{\rm 2D}$, measured from
    each projection through the clusters. Arrows indicate the median
    of the distributions. Cluster lenses have two dimensional concentrations
    which are $34\%$ higher than the typical cluster in the Universe. 
    \label{fig:conc2d}}
\end{figure*}

Because strong lensing probes the mass in projection, the three
dimensional concentration considered above is not an observable. 
Rather, the two-dimensional concentration is measured from
detailed modeling of the arcs and image positions. For a given 3-D
concentration, we expect an additional bias to
exist in the distribution of 2-D concentrations since lensing
clusters are triaxial and, as discussed above, tend to be viewed down
their major axes.  The left panel of Figure~\ref{fig:conc2d} indicates that
this is indeed the case. There we show the lensing and total
distributions of $c_{2D}\slash c_{3D}$, which is the two-dimensional
concentration (for a given projection) normalized by the 
three-dimensional concentration of each cluster. The histogram is computed
using the cross section for each of the 13,594 unique projections of
the 878 clusters in our simulation volume. Note that median value of
the total distribution deviates from unity by 6\%. This bias relative
to the three dimensional concentrations is caused by the fact that our
two dimensional NFW profile fits underestimate the concentrations
because we fit to the projection of a profile which extends to
infinity (see discussion in \S~\ref{sec:NFW}). Nevertheless, we see
that the lensing distribution of $c_{2D}\slash c_{3D}$ is biased by
$\sim 19$\% relative to the total distribution.

We compare the lensing and total distributions of the two dimensional
concentrations in the right panel of Figure~\ref{fig:conc2d}. The
concentrations of lensing clusters are 34\% larger than than those of
the total cluster population. This bias is caused by the combination
of two effects.  First, strong lensing clusters have three dimensional
concentrations which are on average 18\% larger than the typical
cluster at the same mass, as is indicated by
Figure~\ref{fig:conc}. Second, given a three dimensional
concentration, Figure~\ref{fig:conc2d} indicates that strong lensing
prefers the projections through clusters which give $\sim 19$\% higher
concentrations.  This 34\% concentration bias should be kept in mind
when comparisons are made between concentrations measured from
modeling observed strong lenses to the mean concentrations measured
from N-body simulations.

The foregoing discussion has bearing on the recent measurements of
anomalously high concentrations from detailed modeling of individual
lensing clusters. From combined weak and strong lensing analyses,
\citet{Kneib03} measured a best fit concentration of $c_{\rm NFW}=22$
for CL0024$+$1654, while \citet{Gavazzi03} measured $c_{\rm NFW}=12$
for MS 2137.3$-$2353 \citep[but see][]{DK03}, and \citet{Broad05b}
measured $c_{\rm vir}=13.7$ for Abell 1689. It is unlikely that the
brightest cluster galaxies in these clusters increase the
concentrations over our expectation for dark matter
alone.\footnote{For CL0024$+$1654 and Abell 1689 the arcs occur at
  large radii $\gtrsim 30\arcsec$, where the baryonic component
  contributes only a small fraction $\lesssim 10\%$ of the total mass
  enclosed by the critical curves \citep[see
    e.g.][]{Broad05a}. Baryons are a larger concern for MS
  2137.3$-$2353 because its arcs are at smaller radii $\lesssim
  15\arcsec$, however \citet{Gavazzi03} found that including a mass
  component associated with the central galaxy did not significantly
  change the resulting concentration.}  Note that \citet{Kneib03} and
\citet{Gavazzi03} use a convention for the NFW concentration, $c_{\rm
  NFW} = r_{200}/r_{\rm s}$, where $r_{200}$ is the radius at which
the average density is 200 times the \emph{critical} density, which
differs from our definition of $c_{\rm vir} = r_{vir}/r_{\rm s}$,
where $r_{\rm vir}$ is the radius where the average density is
$\Delta_{\rm vir}(z)$ times the \emph{mean} density. Converting these
concentrations to our convention \citep{White01,HK03}, gives $c_{\rm
  vir}=26.2$ and $c_{\rm vir}=14.6$, for CL0024$+$1654 and MS
2137.3$-$2353, respectively.

These high concentrations are very puzzling, considering that the
distribution in the right panel of Figure~\ref{fig:conc2d} predicts
that the probability for $c_{\rm 2D} > 14$ is less than 2$\%$. Why
should the three best studied lensing clusters in the Universe all
have concentrations on the tail of the concentration distribution?
One possible explanation is that these clusters have been studied
intensively because they are known to contain multiple giant arcs, and
one might expect clusters with multiple high-surface brightness arcs
to be even more biased than the general population of cluster lenses.
While not implausible, \citet{Shirley04} found that $\sim 40-50\%$ of
massive clusters that produce giant arcs show multiple arcs, which
suggests that multi-arc clusters are relatively common.  A definitive
solution to this puzzle of high observed concentrations will clearly
require a large, homogeneously selected statistical sample of strong
lensing clusters.

\begin{figure*}
  \centerline{
    \epsfig{file=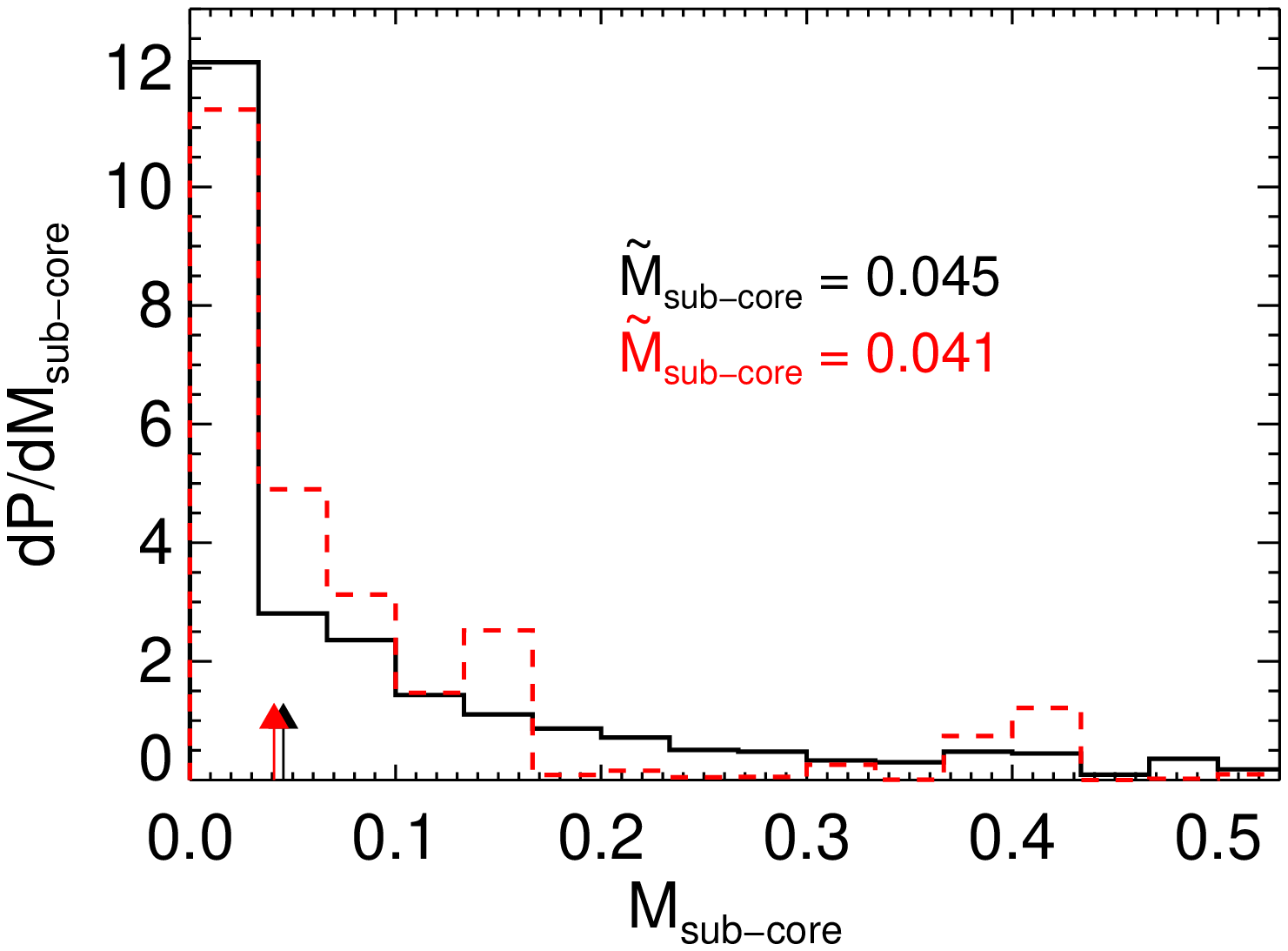,bb=50 0 504 360,width=0.33\textwidth}
    \epsfig{file=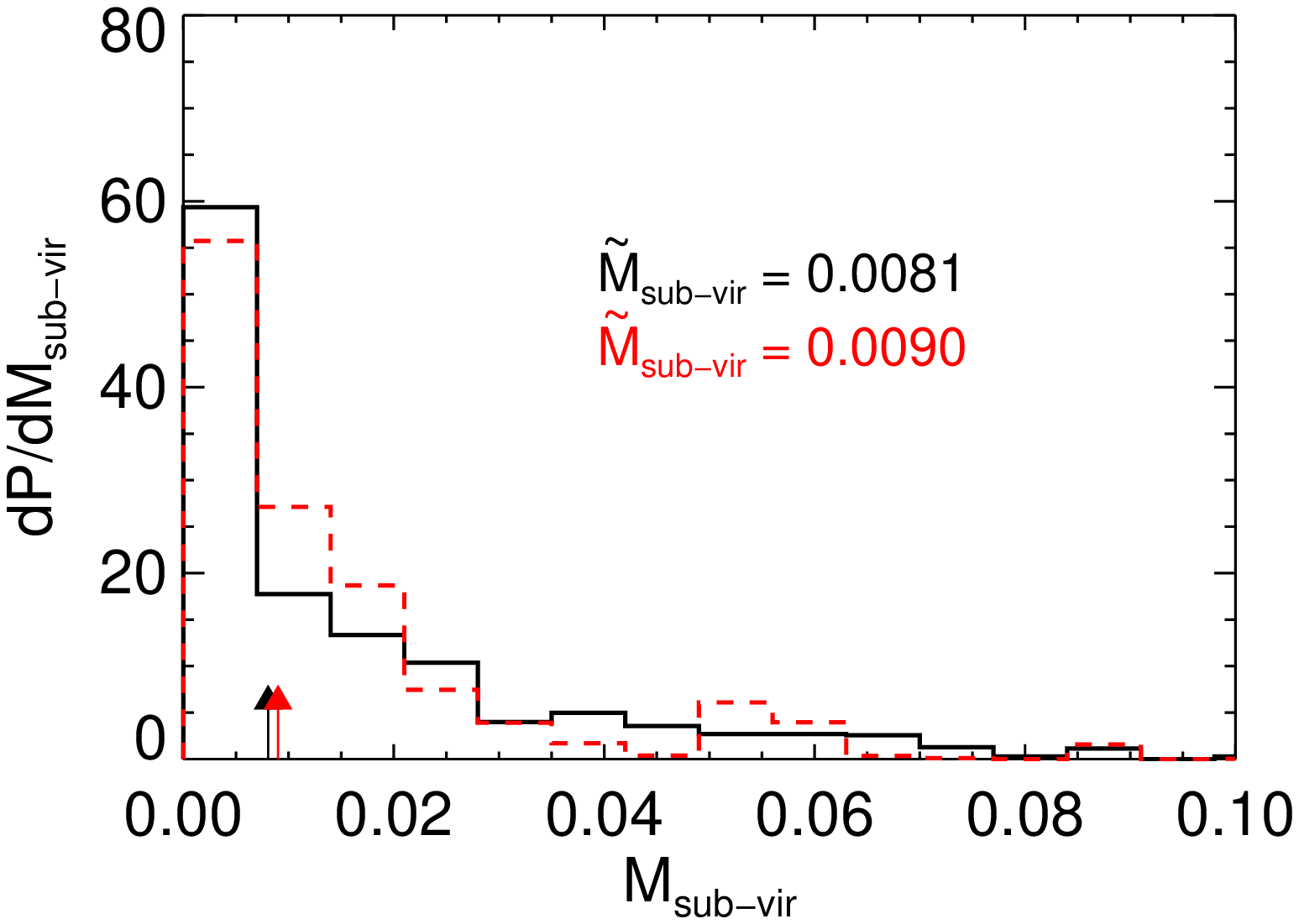,bb=50 0 504 360,width=0.33\textwidth}
    \epsfig{file=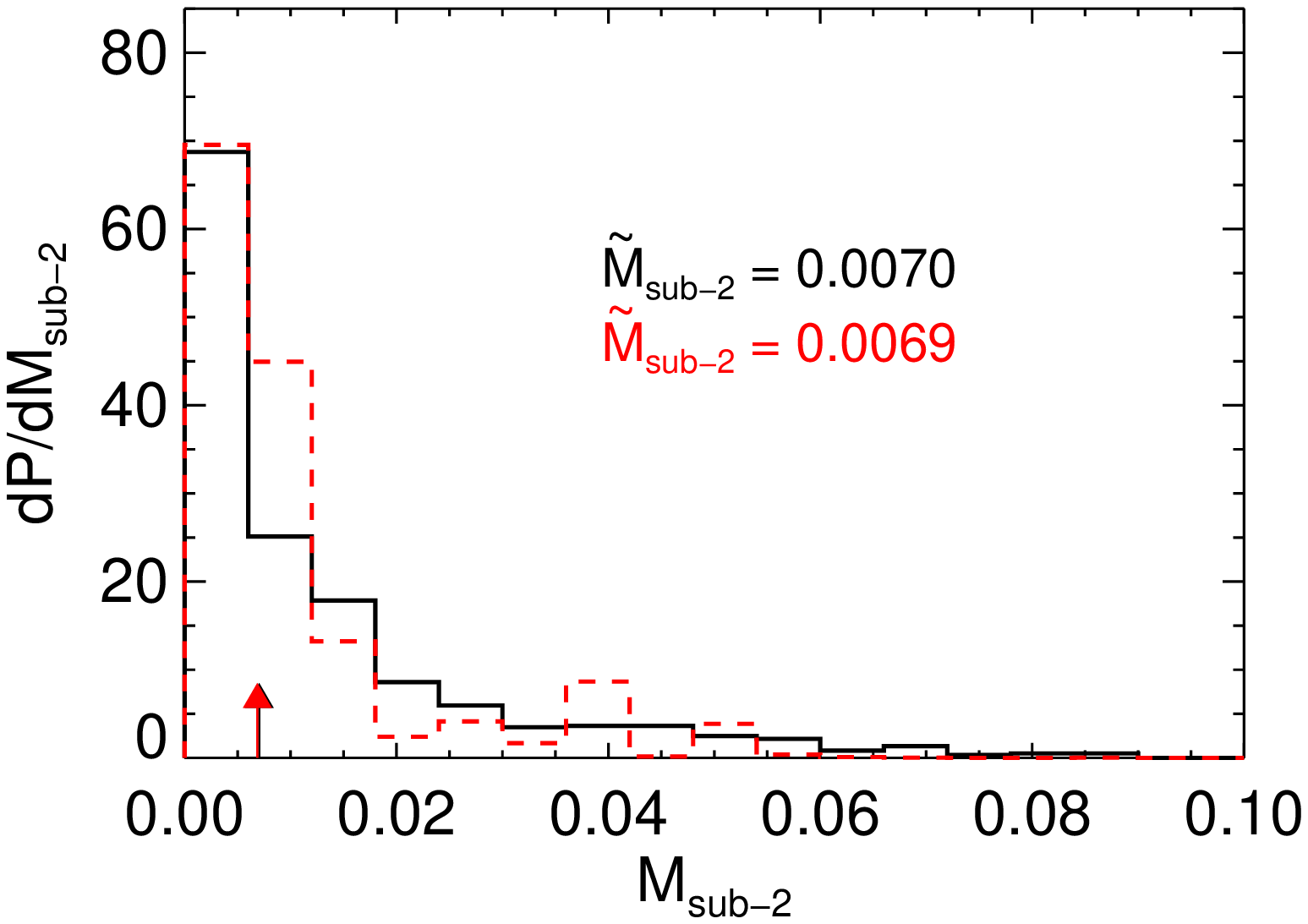,bb=50 0 504 360,width=0.33\textwidth}}
  \caption{ Lensing (dashed) and total (solid) distributions of the
    substructure statistics. The left panel shows the distributions of
    $M_{\rm sub-core}$, which is the mass in substructure normalized
    by the dense core mass. The middle panel shows $M_{\rm sub-vir}$,
    which is the mass in substructure normalized by the virial mass.
    The right panel shows $M_{\rm sub-2}$, which is ratio of the 
    the single most massive substructure to the virial mass. 
    The median of both the lensing and total distributions, indicated by
    the arrows, are labeled on each plot. The agreement between the lensing
    and total distributions for these three different substructure statistics
    clearly indicates that \emph{the population of cluster lenses are no more
      relaxed or disturbed than typical clusters in the Universe.}
    \label{fig:msub}
  }
\end{figure*}


\subsection{Substructure}

In \S~\ref{sec:sub}, we defined three simple statistics which quantify
the amount of substructure in a cluster. The first, $M_{\rm
  sub-core}$, is the ratio of the mass in substructures to the core
mass of the cluster; the second, $M_{\rm sub-vir}$, is the ratio of
the mass in substructure to the virial mass of the cluster; and the
third, $M_{\rm 2-vir}$, is the ratio of the most massive substructure
(i.e. second to the the cluster core) to the virial mass.  The lensing
and total distributions of these three statistics are shown in
Figure~\ref{fig:msub}.


A comparison of the lensing and total distributions of these
statistics indicates that any correlation between strong lensing cross
section and the presence of substructure is marginal at best. First,
focusing attention on Figure~\ref{fig:msub}, which is for
substructures within the virial radius, the median of the statistic
$M_{\rm sub-core}$ for the lensing and total distributions suggests
that lensing clusters have $\sim 10\%$ \emph{less} substructure than
the total cluster population.  The mass in substructure normalized by
the virial mass, $M_{\rm sub-vir}$, is biased in the opposite
direction by $\sim 10\%$ compared to the total cluster
population. Both of these statistics consider all substructures $>
10^{12}~\hmsol$ (see eqn.~\ref{eqn:msub}) and effectively integrate
over the subhalo mass function. In contrast, the statistic $M_{\rm
  sub-2}$ quantifies the binarity of the cluster and is expected to be
largest for a cluster undergoing a major merger. The median value of
$M_{\rm sub-2}$ for lensing selected clusters is indistinguishable
from the total cluster population. 


The fact that $M_{\rm sub-2}$ does not correlate with strong lensing
brings into question the notion that unrelaxed clusters undergoing
major mergers are more effective gravitational lenses.  In particular,
\citet{Torri04} analyzed two different projections through a single
temporally resolved cluster merger and argued that mergers
significantly enhance strong lensing cross sections. While individual
projections can be enhanced by the presence of significant
substructure and similar examples exist for our clusters, these
enhancements due to chance alignments of substructures are diluted, in
the orientation average, by the other projections for which
substructures are not aligned.  In addition, \citet{ZB03} found that
the degree of substructure in dark matter halos anti-correlates with
concentration, and we have seen that more centrally concentrated
clusters tend to dominate the lensing cross section.  Finally, our
conclusion here that substructure does not correlate with strong
lensing is reinforced by the results of the no substructure analog
halos in \S~\ref{sec:analog}. There we saw that smoothly
redistributing all the mass in substructure had a marginal effect,
$\sim 5-10\%$, on the total number of giant arcs produced.  Our
statistical analysis of 13,594 different orientations of 878 clusters
clearly indicates that \emph{the population of cluster lenses are no
  more relaxed or disturbed than typical clusters in the Universe.}

\section{Do the Arcs Discovered in the RCS Agree with $\Lambda$CDM?}
\label{sec:RCS}

\begin{figure}[t!]
  \centerline{
    \epsfig{file=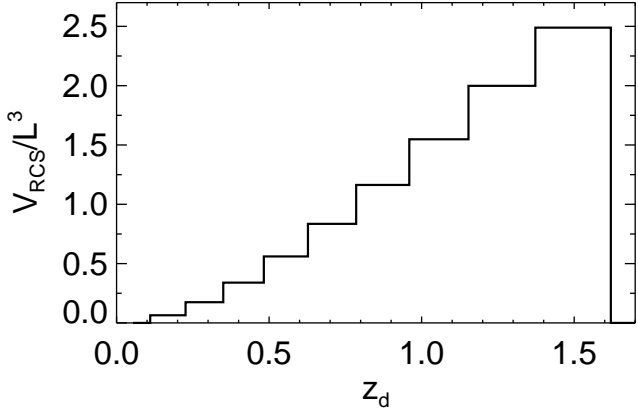,bb=10 10 445 300,width=0.50\textwidth}}
  \caption{ The ratio of volume of the 90 deg$^2$ RCS survey to our
    simulation volume as a function of redshift. The quantity
    $(V_j/L^3)$ is plotted for each snapshot redshift $z_j$ where
    $V_j$ is given by eqn.~(\ref{eqn:Vj}) and $L=320~\hMpc$ is the
    size of our simulation cube.  The snapshot at redshift $z_j$, was
    taken to represent the volume of the Universe over the redshift
    range $[\frac{z_{j} + z_{j-1}}{2},\frac{z_j + z_{j+1}}{2}]$, which
    is represented by the bins of the histogram.
    \label{fig:vol_ratio}
  }
\end{figure}

\begin{figure*}[t!]
  \centerline{
    \epsfig{file=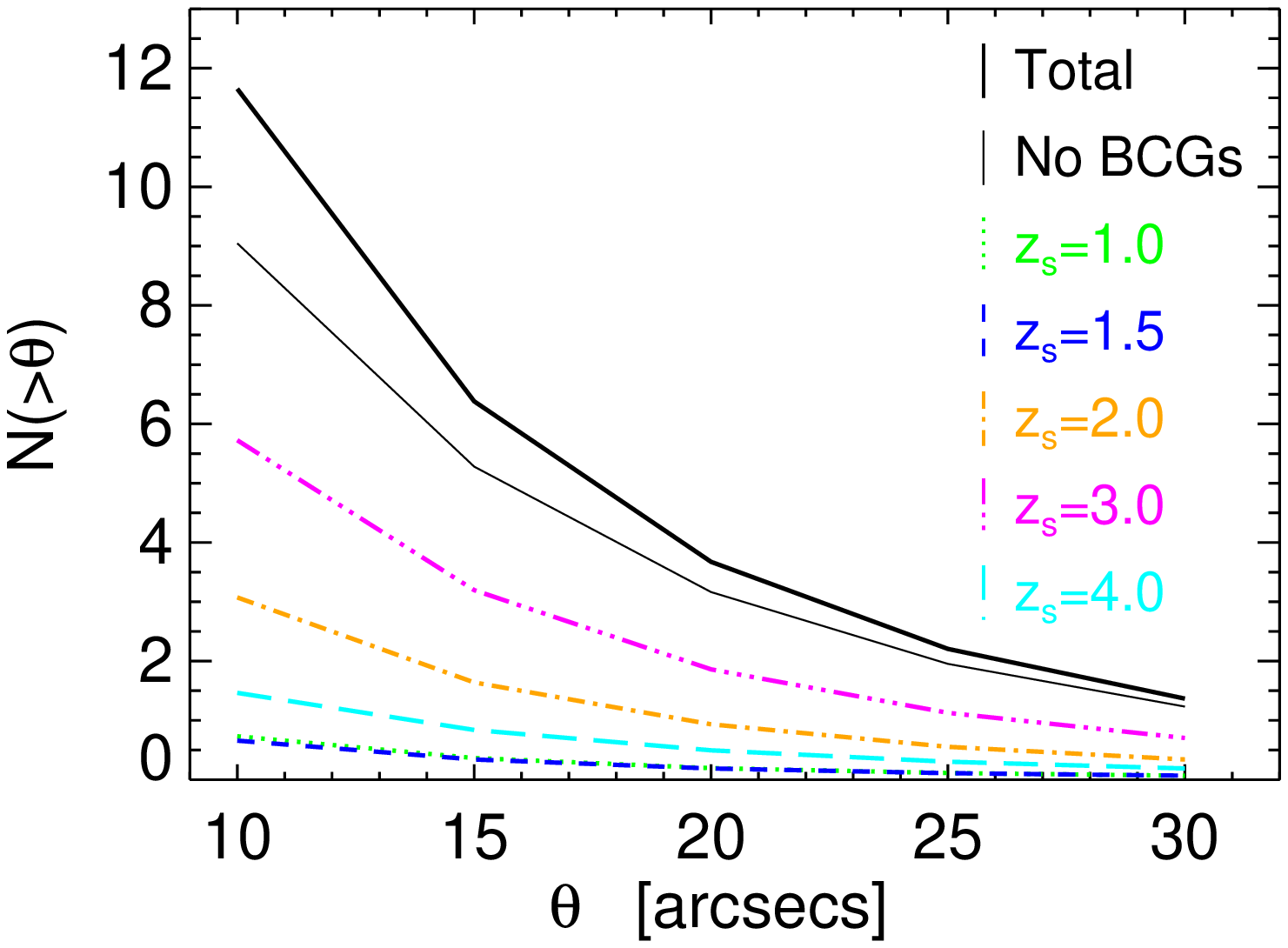,bb=40 0 504 360,width=0.50\textwidth} 
    \epsfig{file=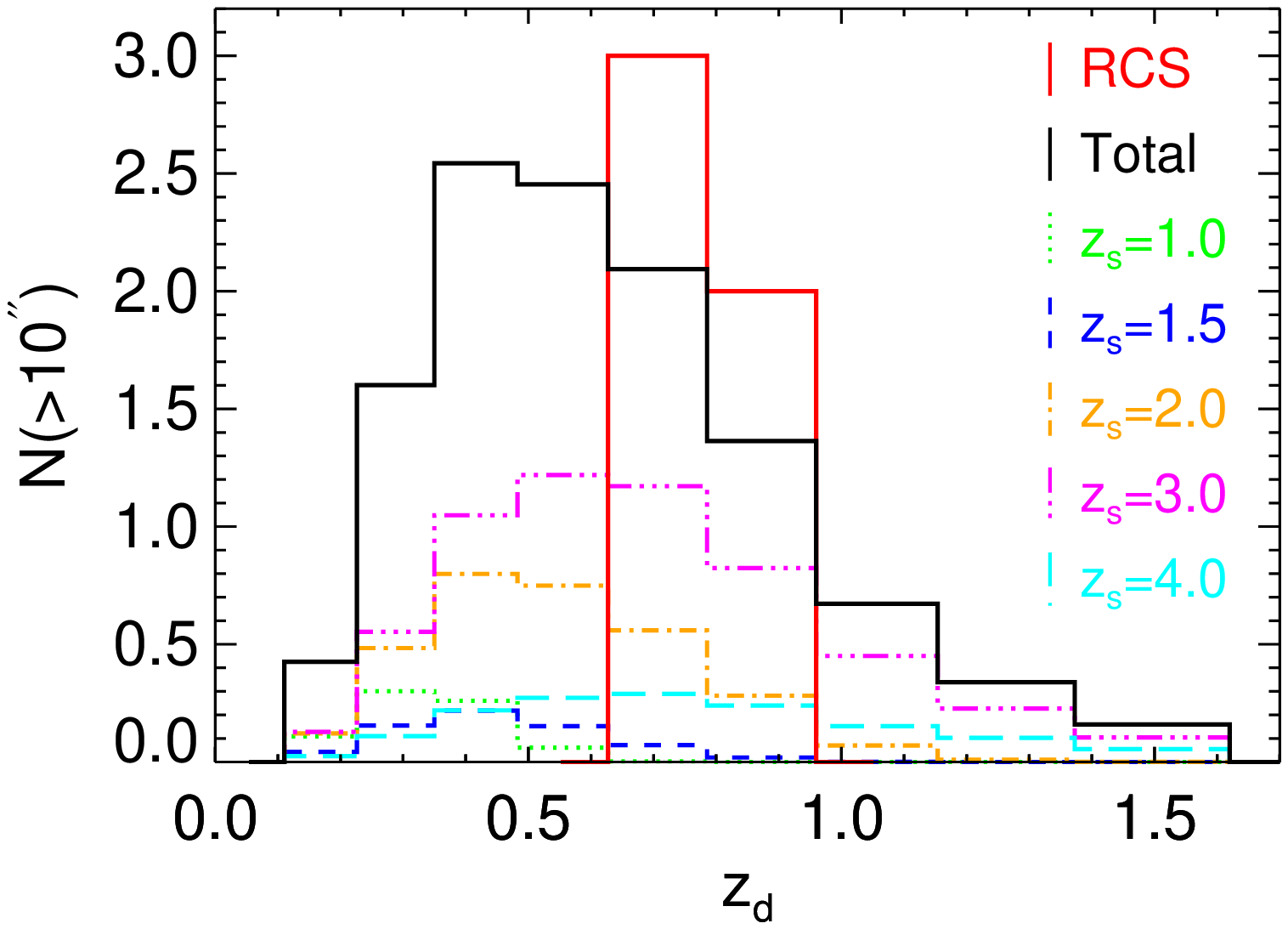,bb=40 0 504 360,width=0.50\textwidth}}
  \caption{ \emph{Left:} Predicted cumulative distribution of giant
    arc separations for the RCS cluster survey. The green (dotted),
    blue (short-dashed), orange (dot-dashed), magenta
    (dot-dot-dashed), and cyan (long dashed) curves are the individual
    contributions from the source planes at $z_{\rm s}=1.0$, $1.5$,
    $2.0$, $3.0$, and $4.0$, respectively. The sum of these curves
    gives the total number of multiply imaged quasars which is the
    thick black (solid) curve. The thin solid (black) line shows the
    total number of lenses if BCGs are neglected and we ray trace
    through dark matter only. \emph{Right:} Redshift histogram of
    cluster lenses with splittings $\theta>10\arcsec$ in the RCS
    survey. The individual contributions from each source plane are
    also plotted with the same line and color scheme as in the left
    panel. The thick black (solid) histogram is the total number of
    giant arcs predicted including BCGs. The red (solid) histogram is
    the observed redshift distribution of cluster lenses in the RCS
    from \citet{Glad03}. \label{fig:RCS}}
\end{figure*}

In this section we revisit the issue of the abundance of giant arcs in
the Red Cluster Sequence cluster survey. \citet{Glad03} surveyed 90
deg$^2$ and found several giant arcs, all in clusters at high redshift
$z>0.6$. DHH were unable to determine whether the number of giant arcs
discovered in the RCS is consistent with the $\Lambda$CDM model
because the simulation cube used in that study was too small
\citep[but see][]{Wamb04a}. Here we revisit this comparison using a
simulation volume $(320/141.3)^3=11.6$ times larger. In
Figure~\ref{fig:vol_ratio} we plot the ratio of the RCS volume to the
volume of our $320~\hMpc$ simulation cube as a function of
redshift. Our simulation volume is larger than the RCS for $z\sim
0.7$, where three of the five RCS arcs which we consider occur (see
below), and comparable to the RCS volume for higher redshifts $0.7
\lesssim z\lesssim 1.0$ where the other two RCS arcs land. Hence, we
have a sufficient volume to sample the very rare massive clusters
likely to be responsible for the RCS strong lensing \citep{Shirley04}. 


\subsection{The Number Density of Background Galaxies}
\label{sec:gals}

\begin{table}[t!]
  \begin{center}
    \caption{Density of Background Sources\label{table:ntt}}
    \begin{tabular}{ccccc}
      \hline
      \hline
      \  $z_{\rm s}$ \  & \ $ z_{\rm min}$ & --  & $z_{\rm max}$ \  & \  $n_{\rm gal}$ \ \\
      \hline
      \  1.0 \  & \ 0.75  & -- &  1.25 \ & \ 6.25 \ \\
      \  1.5 \  & \ 1.25  & -- &  1.75 \ & \ 1.67 \ \\
      \  2.0 \  & \ 1.75  & -- &  2.50 \ & \ 3.96 \ \\
      \  3.0 \  & \ 2.50  & -- &  3.50 \ & \ 3.54 \ \\
      \  4.0 \  & \ 3.50  & -- &  5.00 \ & \ 0.62 \ \\
      \hline
    \end{tabular}
  \end{center}
  \footnotesize NOTES---Density of background sources used in this paper.
\end{table}

Gravitational lensing conserves surface brightness but does not
conserve total integrated flux. Because arcs are resolved, at least in
the tangential direction, arc detection should be limited by surface
brightness rather than integrated flux. In the regime where arcs are
resolved both tangentially and radially, there is no magnification
bias.  However, high redshift galaxies are known to be compact with
half light radii $r_{\rm h}\sim 0.3^{\arcsec}$ \citep{Ferg04} and thus
giant arcs could be radially unresolved. This depends on both the
radial magnification of the lensed image and the resolution or seeing.
About half of the arcs published by \citet{Glad03} are resolved
radially, while the other half are not. For these radially unresolved
arcs there is a small amount of magnification bias because the image
is integrated over the seeing disk in the radial direction.  Because
the typical radial magnifications will be small ($\sim {\rm
  seeing}/r_{\rm h}\sim 2$), we chose to neglect this effect here as it
will be negligible compared to our dominant uncertainty, which turns
out to be the counts of background galaxies. A more careful analysis
could include this radial magnification bias by assuming a typical
seeing and including a distribution of source sizes in the Monte Carlo
part of the ray tracing simulation \citep[see e.g.][]{Shirley04}.

\citet{Glad03} impose a surface brightness cut on their arc sample of
$\mu_{R_C}<24$.  Thus, in order to compare statistics, we require the
surface brightness function of high redshift galaxies, rather than the
luminosity function. Note that because the size distribution of
galaxies is broad \citep{Ferg04} and the luminosities and sizes of
high redshift galaxies are correlated \citep{Bouw04,Truj04}, simply
using the mean size to convert a limiting surface brightness to a
limiting magnitude will not give the correct number counts.

\citet{Fontana00} publish a photometric redshift catalog which
includes the New Technology Telescope (NTT) Deep Field.
\citet{Poli99} measured the half light radii of all the galaxies in
the NTT deep field using image deconvolution techniques. The area of
the NTT deep field is only 4.8 arcmin$^2$. Even for the relatively
coarse redshift bins we use here $\Delta z\sim 1.0$, cosmic variance
for such a small field can be as large as $\sim 50\%$
\citep{Somer04}. Furthermore, half light radii measured from space
based imaging would be clearly preferable to deconvolved ground based
data. However, no wide field space based catalog which includes
redshifts and galaxy sizes exists at the time of writing. Thus, we use
the \citet{Poli99} sizes with the \citet{Fontana00} photo-z's to
determine the number of background galaxies in each redshift bin with
$\mu_{R}<24$. The density of background sources used for each source
redshift bin are listed in Table~\ref{table:ntt}.  We caution that
uncertainty in the density of sources is likely to be a significant
source of error in our comparison of the abundance of giant arcs
to theory. For example, the NTT deep field contains 23 galaxies in the
redshift range [2.5,3.5] with $R<25$, or 4.8 galaxies per
arcmin$^{-2}$.  Yet the incompleteness corrected luminosity function
derived from the $z\sim 3$ Lyman break galaxy sample of
\citet{Steidel99} predicts 1.9 galaxies per arcmin$^{-2}$ in this
range.


\subsection{Comparison with RCS}

The cumulative distribution of image splittings predicted for the RCS
survey from our ray tracing simulations with BCGs added
is shown in the left panel of Figure~\ref{fig:RCS}.
Contributions from the
individual source planes are also shown, along with the total number
predicted from dark matter alone. BCGs significantly increase the
total number of arcs for separations $\theta \lesssim 15\arcsec$
consistent with previous results \citep[DHH;][]{Mene03b,Shirley04}


We restrict our comparison to giant arcs with
separations $\theta > 10\arcsec$, since for smaller separations, our
results will be very sensitive to the details of how we paint BCGs onto
dark halos. We also restrict attention to arcs with length to width
ratios $L/W \gtrsim 10$.  Five of the arcs from Table~1 of
\citet{Glad03} satisfy these criteria. These are the two arcs in
RCS~0224.5-0002 ($z=0.77$), one arc in RCS~1324.5+28245 ($z=0.85$),
one arc in RCS~1419.2+5326 ($z=0.64$), and one of the arcs in the
secondary high redshift sample RCS~2319.9+0038 ($z=0.91$; Gladders
private communication January 2005).

The right panel of Figure~\ref{fig:RCS} compares the predicted
redshift histogram of arcs with $\theta > 10\arcsec$ to the
redshift histogram of the RCS lensing clusters using the same
binning. It is rather surprising that the RCS did not detect any arcs
with $z<0.63$, as the simulations predict $\sim 5$ arcs in this
redshift range. For redshifts $z>0.63$, the simulations predict $\sim
4$ arcs, whereas all five of the RCS arcs are in this range. Although
it is odd that the RCS arcs all pile up between $z=0.6-0.9$, this
anomaly is not statistically significant considering the
small numbers of objects.  We conclude that, at present, there is no
significant excess of high redshift lensing clusters in the RCS
survey. 

Before we conclude this section, we compare the predicted number of
arcs here to the previous results in DHH. They predicted a total of
$\sim 3$ arcs in the RCS for $z<1.0$, whereas here we predict $\sim 8$
arcs in this redshift range. First, the cosmological parameters for
the simulations used in that study were slightly different, most
importantly the simulation used by DHH had $\sigma_8=0.9$, whereas
that used in this study used $\sigma_8=0.95$.  This changes the
abundance $N(>M)$ of massive clusters by $\sim 35-40\%$ for
$M=3\times10^{14}~\hmsol$ and as much as $\sim 170\%$ for $M=3\times
10^{15}~\hmsol$ at $z\sim 0.5$. Although significant, the change in
cluster abundance does not account for the entire disagreement.
Another difference which could be important for the higher redshift
($z\gtrsim 0.6$) clusters is the additional high redshift source
planes ($z_{\rm s}=3.0$ and $z_{\rm s}=4.0$) used here, whereas the
highest redshift source plane considered by DHH was $z_{\rm
  s}=2.0$. However, it is likely that the the factor of $11.6$ smaller
volume used in that study is responsible for most of the
discrepancy. As indicated by Figure~\ref{fig:scatter}, the mass
function is a very steep function of cluster mass and furthermore the
scatter about this mean relation is large because of the large
underlying scatter in cluster properties
(i.e. Figure~\ref{fig:conc}). Large cosmological volumes are thus
required before convergence to the cosmic mean can be achieved.

Have the results of this study converged to the correct cosmic mean
lensing cross sections? First note that the small volume simulated is
at least partly compensated by averaging over a large number of
orientations for each cluster. Although the lensing distribution of
cluster masses in the right panel of Figure~\ref{fig:scatter}
indicates that halos $M\sim 5\times 10^{14}~\hmsol$ are dominating the
total lensing cross section, the tail of this distribution suggests
that we may not yet have converged at the high mass end.  Possible
evidence for a lack of convergence is the fact that \citet{Li05} find
much lower optical depths for a simulation volume of similar size.
Besides convergence to the mean, another serious issue is cosmic
variance. Since the volume we simulated is comparable to the volume of
the RCS, how likely are these two realizations to give the same
answer? Answering this question requires one to measure the highly
non-Gaussian probability distributions of very rare events. This is
clearly beyond the scope of the present study, yet it is a question
which must be tackled if giant arc statistics are to become a
quantitative tool for cosmology.

\section{Summary and Conclusions}
\label{sec:conc}

We attempted to isolate which properties of CDM clusters make them
effective gravitational lenses, by introducing several different types
of `Analog Halos,' which retain one or more of the properties of the
real simulated clusters. The results of this analysis are:

\begin{itemize}

\item[--] Spherical halos underpredict the abundance of giant arcs by
  a factor as large as 50 and triaxial models fall short by a factor
  as much as $60\%$ (see Table~\ref{tab:analog}).

\item[--] Triaxiality increases the number of giant arcs by a factor
  of 4-25 compared to the spherically symmetric halos, because the
  shallow density cusps $\rho \propto r^{-1}$ of CDM halos result in
  an extreme sensitivity to triaxiality \citep[DHH;][]{DK03,BM04}.

\item[--] Projections of halo substructure onto small radii and the
  large scale mass distribution of clusters do not significantly
  influence strong lensing cross sections. The number of giant arcs
  produced by CDM halos is primarily determined by the mass
  distribution with a mean overdensity of $\sim 10,000$.

\item[--] The clumpy cores of dark matter halos result in $\sim
  25-60\%$ more giant arcs than smooth ellipsoids, which suggests that
  the abundance of giant arcs can probe the small scale morphology of
  dark matter halos.

\end{itemize}

We measured the properties of a large ensemble of clusters and
characterized the cluster lens population by computing cross sections
for each cluster with ray tracing simulations.  Our statistical
comparison of the properties of the lensing and total
cluster populations yielded the following results:

\begin{itemize}

\item[--] NFW profiles provide just as good a fit to lensing clusters
  as they do to the total population of clusters.

\item[--] The typical mass of a lensing cluster is $M_{\rm
  vir}=4.5\times 10^{14}~\hmsol$.

\item[--] Lensing clusters have $34\%$ higher concentrations than the
  typical cluster at the same redshift with a similar mass. This bias
  is result of a combination of two effects. First, the lensing
  population is biased towards clusters with higher three dimensional
  concentrations. Second, given a three dimensional concentration, 
  orientation bias will favor projections along the major axis 
  with higher two dimensional concentrations. 
  
\item[--] The anomalously high concentrations $c > 14$ recently
  reported by several groups \citep{Kneib03,Gavazzi03,Broad05b} appear
  inconsistent with the concentration distribution in our simulations,
  which predict that $< 2\%$ of lensing clusters should have concentrations
  this high.

\item[--] The population of cluster lenses are no more relaxed or
  disturbed than typical clusters in the Universe.
  
\item[--] Strong lensing clusters tend to have their principal axis
  aligned with the the line of sight.  The median angle is
  $\left|\cos\theta\right|=0.67$.
  
\item[--] The distribution of axis ratios of strong lensing clusters
  is indistinguishable from the total cluster population. 

\end{itemize}

We revisited the question of whether there is an excess of
giant arcs detected for high redshift clusters in the RCS survey
\citep{Glad03}.  Our simulations predict 9 total arcs in the RCS
survey, whereas five were discovered.  At low redshift ($z\lesssim 0.6$)
the RCS found zero arcs, which is discrepant with our prediction of
$\sim 5$. At high redshift ($z\gtrsim 0.6$), our prediction of $\sim 4$
giant arcs is consistent with the five discovered in the RCS. There is
no significant excess of high redshift lensing clusters in the RCS
survey over predictions from the $\Lambda$CDM model. 

Finally, we emphasize that the results in this work were all based on
dissipationless dark matter only N-body simulations.  Thus, we
implicitly assumed that dissipative baryonic processes such as
heating, cooling, turbulence, and star formation have a negligible
effect on the average surface mass density of a cluster within
$r_{10,000}\sim 300~\hkpc$ (although we did account for the effect of
brightest cluster galaxies in \S~\ref{sec:BCG}) .  While
hydrodynamical simulations of cluster formation are making rapid
progress \citep{Gnedin04,Kazan04,Krav05}, they are not yet mature
enough to make definitive predictions about the effects on the 
mass distribution on scales relevant for strong lensing \citep{Puch05}.  
This is a fruitful topic for future research, considering the 
potentially significant impact these processes could have on our 
interpretation of strong lensing in clusters.

\acknowledgments 

We acknowledge helpful discussions with Olivier Dore, Shirley Ho,
David Hogg, Mike Gladders, Peter Schneider, Risa Wechsler, and Joachim
Wambsganss.  Jeffrey Newman and Alice Shapley helped us with the
number counts of galaxies, and Yeong-Shang Loh pointed us to relevant
references on BCGs. We thank Joanne Cohn and Martin White for reading
an early version of this manuscript and providing helpful
comments. JFH would like to thank his thesis advisors David Spergel
and Michael Strauss for advice and guidance during his time in
Princeton, where this work was begun.  For part of this study JFH was
supported by Proctor Graduate fellowship at Princeton University and
by a generous gift from the Paul \& Daisy Soros Fellowship for New
Americans. The program is not responsible for the views expressed.
JFH and ND are supported by NASA through Hubble Fellowship grants \#
01172.01-A and 01148.01-A respectively, awarded by the Space Telescope
Science Institute, which is operated by the Association of
Universities for Research in Astronomy, Inc., for NASA, under contract
NAS 5-26555.  Computer time was provided by the National Computational
Science Alliance under program \#MCA04N002P, and some computations were
performed on the NSF Terascale Computing System at the Pittsburgh
Supercomputing Center. The ray tracing simulations used computational
facilities at Princeton supported by NSF grant AST-0216105.

 


\begin{thebibliography}
\frenchspacing

\bibitem[Allgood et al.(2005)]{Allgood05} Allgood, B \etal 2005, in
  preparation

\bibitem[Arabadjis, Bautz, \& Garmire(2002)]{ABG02} 
Arabadjis, J.~S., Bautz, M.~W., \& Garmire, G.~P.\ 2002, \apj, 572, 66

\bibitem[Barnes \& Efstathiou(1987)]{BE87} Barnes, J.~\& 
Efstathiou, G.\ 1987, \apj, 319, 575


\bibitem[Bartelmann \& Weiss(1994)]{Bart94} Bartelmann, M.~\& 
Weiss, A.\ 1994, \aap, 287, 1

\bibitem[Bartelmann, Steinmetz, \& Weiss(1995)]{BSW95}
Bartelmann, M., Steinmetz, M., \& Weiss, A.\ 1995, \aap, 297, 1 

\bibitem[Bartelmann(1995)]{Bart95} Bartelmann, M.\ 1995, \aap, 
303, 643 

\bibitem[Bartelmann(1996)]{Bart96} Bartelmann, M.\ 1996, \aap, 
313, 697

\bibitem[Bartelmann et al.(1998)]{Bart98} Bartelmann, M., 
Huss, A., Colberg, J.~M., Jenkins, A., \& Pearce, F.~R.\ 1998, \aap, 330, 1 

\bibitem[Bartelmann et al.(2003)]{Bart03} Bartelmann, M., 
Meneghetti, M., Perrotta, F., Baccigalupi, C., \& Moscardini, L.\ 2003, 
\aap, 409, 449 

\bibitem[Bartelmann \& Meneghetti(2004)]{BM04} Bartelmann, 
M.~\& Meneghetti, M.\ 2004, \aap, 418, 413 

\bibitem[Bode \& Ostriker(2003)]{BO03} Bode, P.~\& Ostriker, 
J.~P.\ 2003, \apjs, 145, 1

\bibitem[Bouwens et al.(2004)]{Bouw04} Bouwens, R.~J., 
Illingworth, G.~D., Blakeslee, J.~P., Broadhurst, T.~J., \& Franx, M.\ 
2004, \apjl, 611, L1 


\bibitem[Broadhurst et al.(2005a)]{Broad05a} Broadhurst, T., et 
al.\ 2005, \apj, 621, 53 

\bibitem[Broadhurst et al.(2005b)]{Broad05b} Broadhurst, T., 
Takada, M., Umetsu, K., Kong, X., Arimoto, N., Chiba, M., \& Futamase, T.\ 
2005, \apjl, 619, L143 

\bibitem[Bryan \& Norman(1998)]{BN98} Bryan, G.~L.~\& 
Norman, M.~L.\ 1998, \apj, 495, 80 

\bibitem[Bullock et al.(2001)]{Bullock01} Bullock, J.~S., Kolatt, 
T.~S., Sigad, Y., Somerville, R.~S., Kravtsov, A.~V., Klypin, A.~A., 
Primack, J.~R., \& Dekel, A.\ 2001, \mnras, 321, 559 

\bibitem[Carlstrom, Holder, \& Reese(2002)]{Carl02} Carlstrom, J.~E., Holder, G.~P., 
\& Reese, E.~D., \araa, 40, 643


\bibitem[Dalal \& Keeton(2003)]{DK03} Dalal, N.~\& Keeton, 
C.~R.\ 2003, ArXiv Astrophysics e-prints, astro-ph/0312072

\bibitem[Dalal, Holder, \& Hennawi(2004)]{DHH} Dalal, N., 
  Holder, G., \& Hennawi, J.~F.\ 2004, \apj, 609, 5

\bibitem[Dalal, Hennawi, \& Bode (2005)]{DHB05} Dalal, N., Hennawi, 
J.~F., \& Bode, P.\ 2005, \apj, 622, 99

\bibitem[Davis \etal(1985)]{Davis85} 
Davis, M., Efstathiou, G., Frenk, C.~S., \& White, S.~D.~M.\ 1985, \apj, 
292, 371

\bibitem[De Lucia et al.(2004)]{DL04} De Lucia, G., 
Kauffmann, G., Springel, V., White, S.~D.~M., Lanzoni, B., Stoehr, F., 
Tormen, G., \& Yoshida, N.\ 2004, \mnras, 348, 333 

\bibitem[Ebeling et al.(2001)]{Ebel01} Ebeling, H., Edge, 
A.~C., \& Henry, J.~P.\ 2001, \apj, 553, 668 

\bibitem[Edge \& Stewart(1991)]{ES91} Edge, A.~C.~\& 
Stewart, G.~C.\ 1991, \mnras, 252, 428

\bibitem[Ettori, Fabian, Allen, \& Johnstone(2002)]{Ettori02}
Ettori, S., Fabian, A.~C., Allen, S.~W., \& Johnstone, R.~M.\ 2002, \mnras, 
331, 635

\bibitem[Faber \& Jackson(1976)]{FJ76} Faber, S.~M.~\& 
Jackson, R.~E.\ 1976, \apj, 204, 668 

\bibitem[Fan et al.(2001)]{Fan01} Fan, X., et al.\ 2001, \aj, 
121, 54

\bibitem[Ferguson et al.(2004)]{Ferg04} Ferguson, H.~C., et 
al.\ 2004, \apjl, 600, L107 

\bibitem[Fisher, Illingworth, \& Franx(1995)]{FIF95} Fisher, 
D., Illingworth, G., \& Franx, M.\ 1995, \apj, 438, 539

\bibitem[{{Flores} {et~al.}(2000){Flores}, {Maller}, \& {Primack}}]{Flores00}
{Flores}, R.~A., {Maller}, A.~H., \& {Primack}, J.~R. 2000, \apj, 535, 55

\bibitem[Fontana et al.(2000)]{Fontana00} Fontana, A., D'Odorico, 
S., Poli, F., Giallongo, E., Arnouts, S., Cristiani, S., Moorwood, A., \& 
Saracco, P.\ 2000, \aj, 120, 2206

\bibitem[Gavazzi et al.(2003)]{Gavazzi03} Gavazzi, R., Fort, B., 
Mellier, Y., Pell{\' o}, R., \& Dantel-Fort, M.\ 2003, \aap, 403, 11 

\bibitem[Gnedin et al.(2004)]{Gnedin04} Gnedin, O.~Y., Kravtsov, 
A.~V., Klypin, A.~A., \& Nagai, D.\ 2004, \apj, 616, 16 

\bibitem[Gill, Knebe, \& Gibson(2004)]{GKG04} Gill, S.~P.~D., 
Knebe, A., \& Gibson, B.~K.\ 2004, ArXiv Astrophysics e-prints, 
astro-ph/0404258

\bibitem[Gladders et al.(2003)]{Glad03} Gladders, M.~D., 
Hoekstra, H., Yee, H.~K.~C., Hall, P.~B., \& Barrientos, L.~F.\ 2003, \apj, 
593, 48 

\bibitem[Gladders \& Yee(2004)]{GY04} Gladders, M.~D., \& 
Yee, H.~K.~C.\ 2004, ArXiv Astrophysics e-prints, astro-ph/0411075

\bibitem[Schulz et al.(2005)]{Schulz05} Schulz, A.~E. \etal 2005, in preparation


\bibitem[Ho \& White(2004)]{Shirley04} Ho, S., \& White, M.\ 
  2004, ArXiv Astrophysics e-prints, astro-ph/0408245
  
\bibitem[Hockney \& Eastwood (1981)]{HoEa81} Hockney, R.~W.~\& 
Eastwood, J.~W.\ 1981, Computer Simulation Using Particles, 
New York: McGraw-Hill, 1981 

\bibitem[Hogg, Baldry, Blanton, \& Eisenstein(2002)]{Hogg02} Hogg,
  D.~W., Baldry, I.~K., Blanton, M.~R., Eisenstein, D.~J. \ 2002,
  ArXiv Astrophysics e-prints, astro-ph/0210394

\bibitem[Hu \& Kravtsov(2003)]{HK03} Hu, W.~\& Kravtsov, 
A.~V.\ 2003, \apj, 584, 702 




\bibitem[Jenkins et al.(2001)]{Jenkins01} Jenkins, A., Frenk, 
C.~S., White, S.~D.~M., Colberg, J.~M., Cole, S., Evrard, A.~E., Couchman, 
H.~M.~P., \& Yoshida, N.\ 2001, \mnras, 321, 372 

\bibitem[Jing \& Suto(2002)]{JS02} Jing, Y.~P.~\& Suto, Y.\ 
2002, \apj, 574, 538 


\bibitem[Kazantzidis et al.(2004)]{Kazan04} Kazantzidis, S., 
Kravtsov, A.~V., Zentner, A.~R., Allgood, B., Nagai, D., \& Moore, B.\ 
2004, \apjl, 611, L73 

\bibitem[Kelson et al.(2002)]{Kelson02} Kelson, D.~D., 
Zabludoff, A.~I., Williams, K.~A., Trager, S.~C., Mulchaey, J.~S., \& 
Bolte, M.\ 2002, \apj, 576, 720

\bibitem[Kneib et al.(2003)]{Kneib03} Kneib, J., et al.\ 2003, 
\apj, 598, 804 


\bibitem[Kochanek et al.(2000)]{Koch00} Kochanek, C.~S., et 
al.\ 2000, \apj, 543, 131 


\bibitem[Kosowsky(2003)]{Koso03} Kosowsky, A.\ 2003, New 
Astronomy Review, 47, 939 

\bibitem[Kravtsov \etal (2003)]{Krav03} Kravtsov A. V., \etal 2003,
ArXiv Astrophysics e-prints, astro-ph/0308519

\bibitem[Kravtsov et al.(2005)]{Krav05} Kravtsov, A.~V., 
Nagai, D., \& Vikhlinin, A.~A.\ 2005, ArXiv Astrophysics e-prints, 
arXiv:astro-ph/0501227

\bibitem[Lacey \& Cole(1994)]{LC94} Lacey, C., \& Cole, S.\ 
1994, \mnras, 271, 676 

\bibitem[Lewis, Buote, \& Stocke(2003)]{LBS03} Lewis, A.~D., 
Buote, D.~A., \& Stocke, J.~T.\ 2003, \apj, 586, 135

\bibitem[Li et al.(2005)]{Li05} Li, G.~L., Mao, S., Jing, 
Y.~P., Bartelmann, M., Kang, X., \& Meneghetti, M.\ 2005, ArXiv 
Astrophysics e-prints, astro-ph/0503172 

\bibitem[Luppino et al.(1999)]{Luppino99} Luppino, G.~A., Gioia, 
I.~M., Hammer, F., Le F{\` e}vre, O., \& Annis, J.~A.\ 1999, \aaps, 136, 
117 







\bibitem[Meneghetti et al.(2000)]{Mene00} Meneghetti, M., 
Bolzonella, M., Bartelmann, M., Moscardini, L., \& Tormen, G.\ 2000, 
\mnras, 314, 338 

\bibitem[Meneghetti, Bartelmann, \& Moscardini(2003a)]{Mene03a} 
Meneghetti, M., Bartelmann, M., \& Moscardini, L.\ 2003, \mnras, 340, 105 

\bibitem[Meneghetti, Bartelmann, \& Moscardini(2003b)]{Mene03b}
Meneghetti, M., Bartelmann, M., \& Moscardini, L.\ 2003, \mnras, 346, 67 


\bibitem[Meneghetti et al.(2004)]{Mene04} Meneghetti, M., 
  Bartelmann, M., Dolag, K., Moscardini, L., Perrotta, F., Baccigalupi, C., 
  \& Tormen, G.\ 2004, ArXiv Astrophysics e-prints, astro-ph/0405070 

\bibitem[Miralda-Escud\'e(1995)]{Miralda95} Miralda-Escude, J.\ 
1995, \apj, 438, 514 

\bibitem[Natarajan \& Kneib(1996)]{NK96} Natarajan, P.~\& 
Kneib, J.\ 1996, \mnras, 283, 1031

\bibitem[Navarro, Frenk, \& White(1997)]{NFW} Navarro, 
J.~F., Frenk, C.~S., \& White, S.~D.~M.\ 1997, \apj, 490, 493

\bibitem[Oegerle \& Hoessel(1991)]{OH91} Oegerle, W.~R.~\& 
Hoessel, J.~G.\ 1991, \apj, 375, 15



\bibitem[Oguri et al.(2003)]{Oguri03} Oguri, M., Lee, J., \& 
  Suto, Y.\ 2003, \apj, 599, 7 
  
  
\bibitem[Oguri et al.(2004)]{Quad04} Oguri, M., et al.\ 2004, 
  \apj, 605, 78 
  
\bibitem[Oguri \& Keeton(2004)]{OK04} Oguri, M.~\& Keeton, 
  C.~R.\ 2004, ArXiv Astrophysics e-prints, astro-ph/0403633

\bibitem[Oguri \& Lee(2004)]{OL04} Oguri, M., \& Lee, J.\ 
2004, \mnras, 355, 120

\bibitem[Phillips et al.(2001)]{Phillips01} Phillips, P.~M., et 
al.\ 2001, \mnras, 328, 1001 

\bibitem[Poli et al.(1999)]{Poli99} Poli, F., Giallongo, E., 
Menci, N., D'Odorico, S., \& Fontana, A.\ 1999, \apj, 527, 662

\bibitem[Power et al.(2003)]{Power03} Power, C., Navarro, 
J.~F., Jenkins, A., Frenk, C.~S., White, S.~D.~M., Springel, V., Stadel, 
J., \& Quinn, T.\ 2003, \mnras, 338, 14 

\bibitem[Puchwein et al.(2005)]{Puch05} Puchwein, E., 
Bartelmann, M., Dolag, K., \& Meneghetti, M.\ 2005, ArXiv Astrophysics 
e-prints, arXiv:astro-ph/0504206 

\bibitem[Richards et al.(2001)]{Richards01} Richards, G.~T., et 
al.\ 2001, \aj, 121, 2308 

\bibitem[Richards et al.(2003)]{Richards03} Richards, G.~T., 
Nichol, R.~C., Gray, A.~G., Brunner, R.~J., Lupton, R.~H., \& Vanden Berk, 
D.~E.\ 2003, American Astronomical Society Meeting, 203

\bibitem[Romer \etal(2001)]{Romer01} Romer, A.~K., Viana, P.~T.~P, Liddle, A.~R.,~\& Mann, R. G. 2001, \apj, 547, 594

\bibitem[Sand et al.(2004)]{Sand04} Sand, D.~J., Treu, T., 
  Smith, G.~P., \& Ellis, R.~S.\ 2004, \apj, 604, 88

\bibitem[Sand et al.(2005)]{Sand05} Sand, D.~J., Treu, T., 
Ellis, R.~S., \& Smith, G.~P.\ 2005, ArXiv Astrophysics e-prints, 
arXiv:astro-ph/0502528 

\bibitem[Schneider et al.(1992)]{sef} Schneider, P., 
Ehlers, J., \& Falco, E.~E.\ 1992, Gravitational Lenses, XIV, 560 pp.~112 
figs..~Springer-Verlag Berlin Heidelberg New York.~ Also Astronomy and 
Astrophysics Library)  

\bibitem[Schwan et al.(2003)]{Schwan03} Schwan, D., et al.\ 
2003, New Astronomy Review, 47, 933 

\bibitem[Sheth(2003)]{Sheth03} Sheth, R.~K.\ 2003, \mnras, 345, 
1200 
  
\bibitem[Sheth \& Jain(2003)]{SJ03} Sheth, R.~K., \& Jain, 
B.\ 2003, \mnras, 345, 529

\bibitem[Smith et al.(2001)]{Smith01} Smith, G.~P., Kneib, J., 
Ebeling, H., Czoske, O., \& Smail, I.\ 2001, \apj, 552, 493 

\bibitem[Somerville et al.(2004)]{Somer04} Somerville, R.~S., 
Lee, K., Ferguson, H.~C., Gardner, J.~P., Moustakas, L.~A., \& Giavalisco, 
M.\ 2004, \apjl, 600, L171 

\bibitem[Spergel et al.(2003)]{Spergel03} Spergel, D.~N., et al.\ 
2003, \apjs, 148, 175 

\bibitem[Steidel et al.(1999)]{Steidel99} Steidel, C.~C., 
  Adelberger, K.~L., Giavalisco, M., Dickinson, M., \& Pettini, M.\ 1999, 
  \apj, 519, 1
  
\bibitem[Torri et al.(2004)]{Torri04} Torri, E., Meneghetti, 
M., Bartelmann, M., Moscardini, L., Rasia, E., \& Tormen, G.\ 2004, \mnras, 
349, 476 

\bibitem[Trujillo et al.(2004)]{Truj04} Trujillo, I., et al.\ 
2004, \apj, 604, 521 

\bibitem[Tyson, Kochanski, \& dell'Antonio(1998)]{Tyson98} 
Tyson, J.~A., Kochanski, G.~P., \& dell'Antonio, I.~P.\ 1998, \apjl, 498, 
L107 

\bibitem[Vanden Berk et al.(2001)]{vanden01} Vanden Berk, D.~E., 
et al.\ 2001, \aj, 122, 549

\bibitem[Warren \etal(1992)]{Warren92} 
Warren, M.~S., Quinn, P.~J., Salmon, J.~K., \& Zurek, W.~H.\ 1992, \apj, 
399, 405 

\bibitem[Wambsganss et al.(2004a)]{Wamb04a} Wambsganss, J., 
Bode, P., \& Ostriker, J.~P.\ 2004, \apjl, 606, L93 

\bibitem[Wambsganss et al.(2004b)]{Wamb04b} Wambsganss, J., 
Bode, P., \& Ostriker, J.~P.\ 2004, ArXiv Astrophysics e-prints, 
astro-ph/0405147 


\bibitem[Wechsler et al.(2002)]{Wech02} Wechsler, R.~H., 
Bullock, J.~S., Primack, J.~R., Kravtsov, A.~V., \& Dekel, A.\ 2002, \apj, 
568, 52 

\bibitem[Weller, Ostriker, \& Bode(2004)]{Weller04} Weller, J., 
Ostriker, J.~P, \& Bode, P.\ 2004, ArXiv Astrophysics e-prints, 
astro-ph/0405445

\bibitem[White(2001)]{White01} White, M.\ 2001, \aap, 367, 27 


\bibitem[Wright \& Brainerd(2000)]{WB00} Wright, C.~O., \& 
  Brainerd, T.~G.\ 2000, \apj, 534, 34 




\bibitem[Zhdanov \& Surdej(2001)]{ZS01} Zhdanov, V.~I.~\& 
Surdej, J.\ 2001, \aap, 372, 1

\bibitem[Zentner \& Bullock(2003)]{ZB03} Zentner, A.~R., \& 
Bullock, J.~S.\ 2003, \apj, 598, 49 

\end{thebibliography}
\end{document}